\documentclass[fleqn,usenatbib]{mnras}

\usepackage{color}
\usepackage[T1]{fontenc}
\usepackage{ae,aecompl}
\usepackage{xcolor}
\usepackage{ulem}

\usepackage{graphicx}	
\usepackage{amsmath}	
\usepackage{cases}
\usepackage{adjustbox}

\usepackage{graphicx}	
\usepackage{amssymb}	
\usepackage{newtxtext,newtxmath}
\usepackage{cases}
\usepackage{array,multirow}
\usepackage{upgreek}
\usepackage{textcomp}
\usepackage{rotating}
\usepackage{amsmath}

\newcommand{\Msol}{\,{\rm M}_{\odot}}
\newcommand{\Rsol}{\,{\rm R}_{\odot}}

\newcommand{\km}{\,\mathrm{km}}

\newcommand{\pc}{\,\mathrm{pc}}
\newcommand{\yr}{\,\mathrm{yr}}
\newcommand{\mstar}{\,M_{\star}}
\newcommand{\rstar}{\,R_{\star}}

\newcommand{\mesa}{{\small MESA}}

\usepackage{tikz}

\newcommand*\fullcirc[1][1ex]{\tikz\fill (0,0) circle (#1);}

\defcitealias{Ryu+2022}{Paper 1}
\defcitealias{Ryu+2023}{Paper 2}
\defcitealias{Ryu+2023b}{Paper 3}

\title[Hydrodynamics of 3-body Encounters]{Close Encounters of Star - Black Hole Binaries with Single Stars}

\author[T. Ryu et al.]{%
Taeho Ryu$^{1,2}$,\thanks{E-mail: tryu@mpa-garching.mpg.de}
Selma E. de Mink$^{1,5}$,
Rob Farmer$^{1}$,
R\"udiger Pakmor$^{1}$,
Rosalba Perna$^{3,4}$,
Volker Springel$^{1}$
\vspace*{0.1cm}\\%
$^{1}$ Max Planck Institute for Astrophysics, Karl-Schwarzschild-Str.~1, 85748 Garching, Germany\\%
$^{2}$ Physics and Astronomy Department, Johns Hopkins University, Baltimore, MD 21218, USA\\%
$^{3}$ Department of Physics and Astronomy, Stony Brook
  University, Stony Brook, NY 11794-3800, USA\\%
$^{4}$ Center for Computational Astrophysics, Flatiron Institute, New York, NY 10010, USA\\%
$^{5}$ Anton Pannekoek Institute for Astronomy, University of Amsterdam, Science Park 904, 1098XH Amsterdam, The Netherlands
}

\date{Accepted XXX. Received YYY; in original form ZZZ}

\pubyear{2022}

\begin{document}
\label{firstpage}
\pagerange{\pageref{firstpage}--\pageref{lastpage}}
\maketitle

\begin{abstract}
Multi-body dynamical interactions of binaries with other objects are one of the main driving mechanisms for the evolution of star clusters. It is thus important to bring our understanding of three-body interactions beyond the commonly employed point-particle approximation. To this end we here investigate the hydrodynamics of three-body encounters between star-black hole (BH) binaries and single stars, focusing on the identification of final outcomes and their long-term evolution and observational properties, using the moving-mesh hydrodynamics code {\small AREPO}. This type of encounters produces five types of outcomes: stellar disruption, stellar collision, weak perturbation of the original binary, binary member exchange, and triple formation. The two decisive parameters are the binary phase angle, which determines which two objects meet at the ﬁrst closest approach, and the impact parameter, which sets the boundary between violent and non-violent interactions. When the impact parameter is smaller than the semimajor axis of the binary, tidal disruptions and star-BH collisions frequently occur when the BH and the incoming star first meet, while the two stars mostly merge when the two stars meet first instead. In both cases, the BHs accrete from an accretion disk at super-Eddington rates, possibly generating flares luminous enough to be observed. The stellar collision products either form a binary with the BH or remain unbound to the BH. Upon collision, the merged stars are hotter and larger than main sequence stars of the same mass at similar age. Even after recovering their thermal equilibrium state, stellar collision products, if isolated, would remain hotter and brighter than main sequence stars until becoming giants.
\end{abstract}



\section{Introduction}\label{sec:intro}

Dynamical interactions between stars and compact objects in dense environments play a fundamental role in a variety of astrophysical settings, from influencing the thermodynamic state of a star cluster \citep{Hut1992}, to altering original planetary architectures (e.g. \citealt{Wang2020,Li2020}), to forming binary black holes (BHs) (e.g. \citealt{Portegieszwart2000,Samsing2014,Perna2019,Rodriguez2015,Antonini2016,Fragione2019,Mapelli2021}) which may make up a significant contribution to the BH-BH mergers detected via gravitational waves (GWs) by the LIGO and Virgo observatories \citep{LIGO2021}.

We are living in an exciting era of transient surveys where the number of transients will exponentially grow soon with detections by ongoing (e.g., the Zwicky Transient Facility \footnote{https://www.ztf.caltech.edu}) and future (e.g., Vera Rubin Observatory\footnote{https://www.lsst.org} and ULTRASAT\footnote{https://www.weizmann.ac.il/ultrasat}) surveys. However, the origin of many transients such as the newly discovered class of the Fast Blue Optical Transients \citep[FBOTs,][]{Drout+2014} remains unknown. For their reliable identification, it is imperative to understand possible mechanisms for the formation of various types of transients \citep[e.g.,][]{Margutti+2019}. In particular, when a dynamical interaction between a star and a BH brings them within a very close distance, the star	can be destroyed via its strong interactions with the BH, leading to the production of a bright flare. Due to the expectation of these transient electromagnetic signatures, the study of close dynamical interactions between stars and compact objects is an especially timely one in light of both ongoing and upcoming transient surveys.

When close interactions	involve	a three-body encounter between a binary and a tertiary object, there is a richer set of possible outcomes compared to the case of a close encounter between a star and a stellar-mass BH where, depending on the closeness of the	encounter, the main outcome is a partial or full disruption of the star \citep{Perets2016,Wang2021TDE}.	While so far few hydrodynamical simulations involving binaries have been carried out \citep[e.g.][]{McMillan+1991,Goodman1991,Lopez+2019}, the impending increase in the number of detectable transients, and at the same time the importance of compact object binaries for	GW observations, make these investigations especially timely. Rarer events, which may have not been detectable to date, may likely be	in the near future.

Since encounters among more than two objects can become chaotic and are not analytically tractable in general, most of the theoretical understanding of dynamical interactions is based on numerical experiments using $N$-body simulations in which the trajectories of stars or BHs, approximated as a point particle, are integrated under the gravitational forces over time \citep[e.g.,][]{Fregeau+2004,Leigh+2016,Leigh+2017,Ryu+2017b,Trani+2019}. Those experiments have provided profound insights into dynamical interactions, particularly the statistical properties of outcomes. In addition, assuming a finite size of the point masses, one can in principle investigate the occurrence rate of transients, such as tidal disruption events or stellar collisions, using $N$-body simulations with finite size of the point masses \citep[e.g.,][]{Fregeau+2004,Ryu+2022c}. However, in these studies, non-linear hydrodynamic effects on stars or on surrounding media, such as tidal deformation or shocks, which are essential for the prediction of observables and an accurate identification of outcomes, are ignored or treated very approximately.

We have	therefore recently began a systematic hydrodynamical investigation of 3-body close encounters between a binary and a tertiary object, involving spatially resolved stars, which we have presented in a series of papers. In \citet[][\citetalias{Ryu+2022} in the following]{Ryu+2022}, we investigated the outcomes of close encounters between main sequence stars and stellar-mass binary BHs, using 3D smoothed particle hydrodynamics simulations, and for a variety of initial orbital parameters and encounter geometries. We found a rich phenomenology for the predicted accretion rates (which can be considered to zeroth order a proxy for the luminosity): while some encounters lead to signatures similar to those typical of the 2-body encounters, other situations carry clear signatures of the binarity, with the accretion rate modulated over the binary period, as both BHs alternate in stripping mass from the star. We further found that the interaction of the BH binary with the star and the stellar disruption itself can produce a significant feedback on the binary orbital parameters, quantitatively differing from the cases of pure scattering \citep{Wang2021}. A single close encounter can produce changes of up of unity in the GW-driven merger timescale.

In \citet[][\citetalias{Ryu+2023} in the following]{Ryu+2023}, we explored the outcomes of close 3-body encounters in which the binary is composed of two main sequence stars, while the incoming tertiary object is a stellar-mass BH, using moving-mesh hydrodynamics  simulations. Again exploring a variety of initial conditions, the simulations uncovered a variety of astrophysical outcomes, from the most standard one of a single star disruption, to a double star disruption, to member exchange leading to the formation of an X-ray binary, to the formation of runaway stars and runaway BHs made active by the accreting debris from the disrupted stars.

\begin{table*}
\begin{tabular}{ c c c c c c c  c c c c c c c} 
\hline
Model number & Model name & \multicolumn{2}{c}{$a$}  &\multicolumn{2}{c}{$b$} & $\phi~[^{\circ}]$  &  $i$  & $t_{\rm p}$ & $P$ & $v_{\rm orb}$\\
\hline
\multicolumn{2}{c}{} & $a_{\rm RL}$ & $\Rsol$ & - & $\Rsol$ & $^{\circ}$ & $^{\circ}$ & hours & days & ${\rm km\,s^{-1}}$ \\
\hline
1 & $a4b2\phi0i30$   & 4 & 67.5 & 2 & 67.5 & 0  & 30 & 39 & 12 & 291 \\
2 & $a4b1\phi0i30$   & 4 & 67.5 & 1  & 33.8 & 0 & 30  & 14  & 12 & 291 \\
3 & $a4b1/2\phi0i30$   & 4 & 67.5 & 1/2  & 16.9 & 0  & 30  & 4.9 & 12 & 291 \\
4 & $a4b1/4\phi0i30$  & 4 & 67.5 & 1/4  & 8.45 & 0  & 30 & 1.7 & 12 & 291 \\
5 & $a4b2\phi180i30$   & 4 & 67.5 & 2 & 67.5  & 180 & 30 & 39 & 12 & 291 \\
6 & $a4b1\phi180i30$   & 4 & 67.5 & 1  & 33.8 & 180 & 30  & 14 & 12 & 291 \\
7 & $a4b1/2\phi180i30$   & 4 & 67.5 & 1/2  & 16.9 & 180 & 30 & 4.9 & 12 & 291 \\
8 & $a4b1/4\phi180i30$  & 4 & 67.5 & 1/4  & 8.45 & 180 & 30   & 1.7 & 12 & 291 \\
\hline
9 & $a4b2\phi0i150$   & 4 & 67.5 & 2  & 67.5   & 0 & 150 & 39 & 12 & 291 \\
10 & $a4b1\phi0i150$   & 4 & 67.5 & 1 & 33.8   & 0  & 150  & 14 & 12 & 291 \\
11 & $a4b1/2\phi0i150$   & 4  & 67.5 & 1/2 & 16.9 & 0  & 150  & 4.9 & 12 & 291 \\
12 & $a4b1/4\phi0i150$   & 4 & 67.5  & 1/4  & 8.45  & 0  & 150 &  1.7 & 12 & 291 \\
13 & $a4b2\phi180i150$   & 4 & 67.5  & 2  & 67.5 & 180  & 150   & 39 & 12 & 291 \\
14 & $a4b1\phi180i150$   & 4 & 67.5  & 1 & 33.8  & 180  & 150 & 14 & 12 & 291 \\
15 & $a4b1/2\phi180i150$  & 4 & 67.5  & 1/2 & 16.9  & 180  & 150   & 4.9 & 12 & 291 \\
16 & $a4b1/4\phi180i150$   & 4 & 67.5  & 1/4 & 8.45  & 180  & 150  &  1.7 & 12 & 291 \\
\hline
17 & $a2b1/2\phi0i30$  & 2  & 33.7  & 1/2  & 8.44 & 0 & 30  & 1.7  & 4.1 & 412\\
18 & $a2b1/2\phi180i30$  & 2  & 33.7 & 1/2  & 8.44 & 180 & 30   & 1.7 & 4.1 & 412\\
19 & $a2b1/2\phi0i150$   & 2 & 33.7 & 1/2   & 8.44 & 0  & 150 &1.7  & 4.1 & 412\\
20 & $a2b1/2\phi180i150$   & 2 & 33.7 & 1/2   & 8.44 & 180  & 150 &1.7 & 4.1 & 412 \\
21 & $a6b1/2\phi0i30$   & 6 & 101 & 1/2  & 25.3 & 0 & 30   & 8.9  & 22 & 238\\
22 & $a6b1/2\phi180i30$   & 6 & 101 & 1/2  & 25.3 & 180 & 30 & 8.9  & 22& 238\\
23 & $a6b1/2\phi0i150$   & 6  & 101 & 1/2  & 25.3  & 0 & 150  & 8.9  & 22& 238\\
24 & $a6b1/2\phi180i150$   & 6 & 101 & 1/2  & 25.3 & 180 & 150  & 8.9  & 22& 238\\
\hline
25 & $a4b1/2\phi0i0$  & 4 & 67.5  & 1/2 & 16.9   & 0 & 0  & 4.9 & 12 & 291 \\
26 & $a4b1/2\phi0i60$   & 4 & 67.5 & 1/2 & 16.9  & 0 & 60 & 4.9 & 12 & 291 \\
27 & $a4b1/2\phi0i120$   & 4 & 67.5 & 1/2  & 16.9  & 0 & 120 & 4.9 & 12 & 291 \\
28 & $a4b1/2\phi0i180$   & 4 & 67.5 & 1/2  & 16.9  & 0 & 180 & 4.9 & 12 & 291 \\
29 & $a4b1/2\phi180i0$  & 4 & 67.5  & 1/2  & 16.9  & 180 & 0 & 4.9 & 12 & 291 \\
30 & $a4b1/2\phi180i60$   & 4 & 67.5  & 1/2 & 16.9  & 180 & 60 & 4.9 & 12 & 291 \\
31 & $a4b1/2\phi180i120$   & 4 & 67.5 & 1/2 & 16.9  & 180 & 120 & 4.9 & 12 & 291 \\
\hline
32 & $a4b1/2\phi45i30$   & 4 & 67.5  & 1/2 & 16.9  & 45  & 30   & 4.9 & 12 & 291 \\
33 & $a4b1/2\phi90i30$   & 4 & 67.5  & 1/2 & 16.9  & 90 & 30  & 4.9 & 12 & 291 \\
34 & $a4b1/2\phi135i30$  & 4 & 67.5  & 1/2 & 16.9  & 135 & 30  & 4.9 & 12 & 291 \\
35 & $a4b1/2\phi225i30$   & 4 & 67.5   & 1/2 & 16.9  & 225  & 30   & 4.9 & 12 & 291 \\
36 & $a4b1/2\phi270i30$   & 4 & 67.5  & 1/2 & 16.9  & 270 & 30  & 4.9 & 12 & 291 \\
37 & $a4b1/2\phi315i30$  & 4 & 67.5  & 1/2 & 16.9  & 315 & 30  & 4.9 & 12 & 291 \\
\hline
\end{tabular}
\caption{The initial model parameters for encounters between a circular binary ($q=0.5$) with total mass of $30\Msol$ and a $10\Msol$ star. The model name (\textit{second} column) conveys the information of key initial parameters explored in this study: for the model names with the format $a(1)b(2)\phi(3)i(4)$, the numerical values encode (1) the initial semimajor axis of the binary $a/a_{\rm RL}$, (2) the impact parameter $b$, (3) the initial phase angle $\phi$ in degrees, and (4) the initial inclination angle $i$ in degrees.  Here, $a_{\rm RL} \simeq 3.12$ and $R_{\star}\simeq 16.9\Rsol$ specify the separation when the star in the binary fills its Roche lobe. The last three columns show the dynamical time $t_{\rm p}$ at pericenter, the orbital period $P$ of the binary, and the relative velocity $v_{\rm orb}$ of the binary members.}\label{tab:initialparameter}
\end{table*}

Most recently, in \citet[][\citetalias{Ryu+2023b} in the following]{Ryu+2023b}, we performed moving-mesh hydrodynamical  simulations of close encounters between single BHs and binaries composed of a main sequence star and a BH. Outcomes were found to range from orbital perturbations of the original binary, to member exchanges either of the BH (hence forming a new X-ray binary), or of the star (hence leading to a binary BH). Deep encounters on the other hand were found to more often lead to the disruption of the star, with accretion rates that can display modulation if the two BHs have bound to form a binary.

In this paper, which is the fourth in our series, we continue our investigation in this area by performing 3D hydrodynamical simulations of close encounters between main sequence, single stars, and binaries composed of a BH and a star, using numerical methods developed in \citetalias{Ryu+2023} and refined in \citetalias{Ryu+2023b}. 
In addition to outcomes already encountered before, we find new astrophysical phenomena, such as the formation of binary star systems via member exchange, and stellar mergers, whose evolution we then follow with the stellar evolution code MESA \citep{Paxton+2011}. 

The paper is organized as follows. Section~\ref{sec:method} describes the numerical methods we use, and the initial conditions of our simulations. Simulation results are presented in Section~\ref{sec:results}, with their astrophysical implications  discussed in Section~\ref{sec:discussion}. We summarize and conclude in Section~\ref{sec:conclusion}.

\begin{figure*}
	\centering
	\includegraphics[width=14cm]{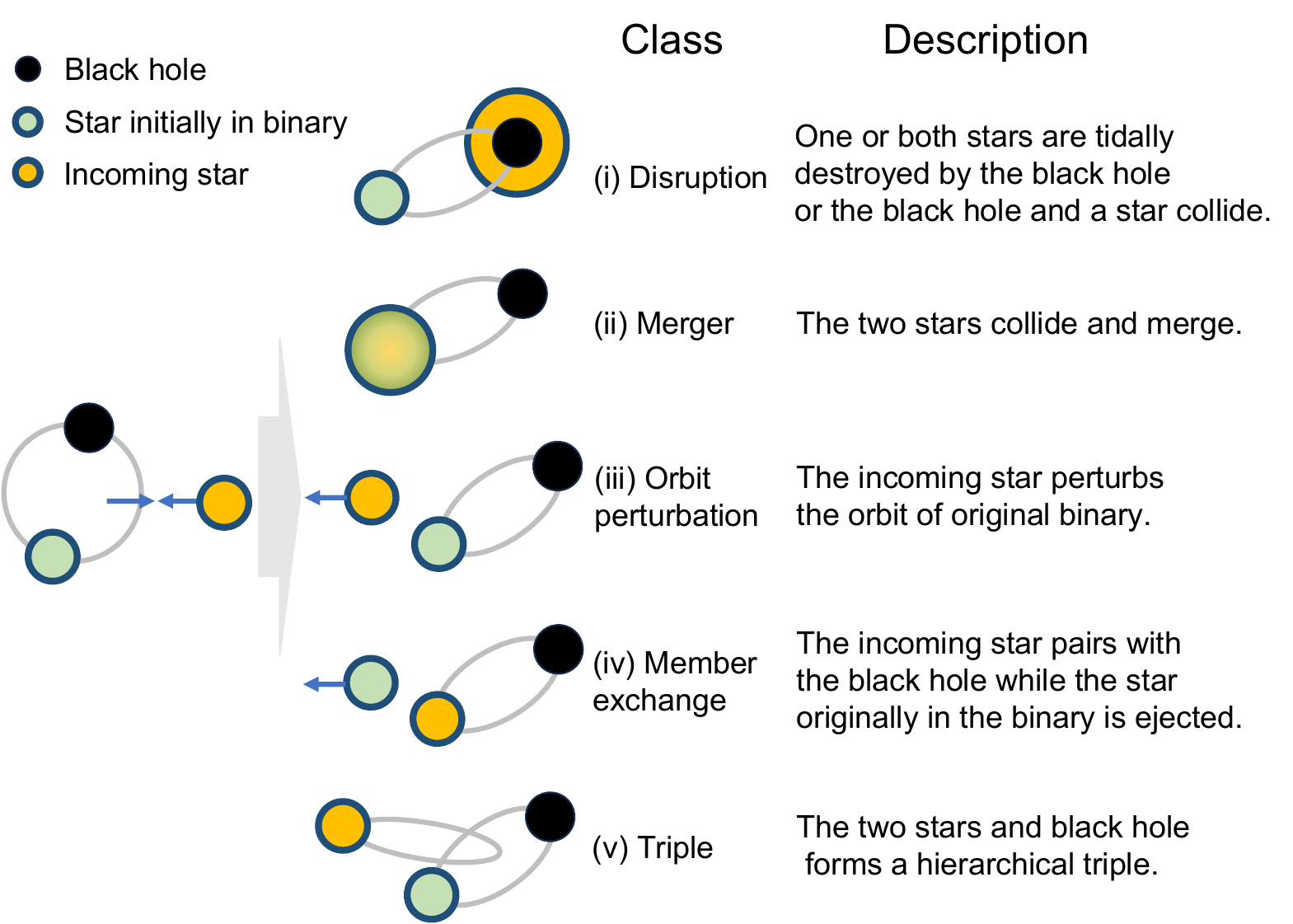}
\caption{Schematic verview of the five outcome classes we identified: \textit{disruption}, \textit{merger}, \textit{orbit perturbation}, \textit{member exchange}, and \textit{triple}, produced in three-body dynamical interactions between BH-star binaries and single stars. See Section~\ref{sub:class} for detailed explanations. }
	\label{fig:summary}
\end{figure*}

\begin{table*}
\resizebox{\textwidth}{!}{
\begin{tabular}{ c c c c c c c | c c c c | c c } 
\hline
Number & Model name & Class  & Disruption? & Collision? & Active BH? & Binary? & Binary type & $a$  & $e$ & $v$ & Single type & $v$\\
\hline
\multicolumn{2}{c}{}  & - & - & - & -  & - & - &  $\Rsol$ & - & km~s$^{-1}$ & - & km~s$^{-1}$ \\
\hline
1 & $a4b2\phi0i30$     & Triple    &   No  &   No  & No   &  Yes &  ($\bigstar-\Diamond)-\fullcirc$  &  76.0 (293)  & 0.79 (0.58) &  - & -  \\
2 &$a4b1\phi0i30$     & Exchange   &   No  &  No   &  No  &  Yes  & $\bigstar-\Diamond$   &  54.4  & 0.366 & 82.4 & $\fullcirc$ & 84.1\\
3 & $a4b1/2\phi0i30$  &  Disruption   &  $\Diamond$   &  No & Yes   & Yes &  $\fullcirc-\bigstar$    &  258  &  0.68  & 46.5 & - & - \\
4 & $a4b1/4\phi0i30$  & Disruption &   $\Diamond$   &  No & Yes   & Yes &  $\fullcirc-\bigstar$    &  947  &  0.918  & 191 & - & - \\
5 & $a4b2\phi180i30$   & Exchange     &  No   &  No   &  No  &   Yes &  $\fullcirc - \Diamond$  &  63.5  & 0.094 &  48.7 & $\bigstar$ & 146 \\
6 & $a4b1\phi180i30$    & Merger  & No & Yes &   Yes & Yes &  $\fullcirc-(\bigstar + \Diamond)$   & 134 & 0.490 & 5.4 & - & -\\
7 & $a4b1/2\phi180i30$  & Merger    &  No   &   Yes  &   Yes &  Yes  &  $\fullcirc-(\bigstar + \Diamond)$  &  129  & 0.56 & 3.83 & - & - \\
8 & $a4b1/4\phi180i30$  & Exchange (I)    & No   &  No   &  Yes  & Yes & $\fullcirc-\Diamond $  &  42.0  &  0.692  & 69.9 & $\bigstar$ & 210 \\
\hline
9 & $a4b2\phi0i150$    & Perturbation    &   No  &   No  &   No & Yes   &  $\fullcirc-\bigstar$    &  62.1  & 0.15 & 59.2 & $\Diamond$ & 230 \\
10 & $a4b1\phi0i150$    & Perturbation     &  No   &   No  &  No  &    Yes &  $\fullcirc-\bigstar$  &  42.4   & 0.719 & 71.2 & $\Diamond$ & 214 \\
11 & $a4b1/2\phi0i150^{\star}$ & Disruption (P)      &  $\Diamond$   &   No  & Yes   & Yes   &  $\fullcirc-\bigstar $  &  -  & - & - & $\Diamond$  & - \\
12 & $a4b1/4\phi0i150$  & Disruption     &    $\Diamond$   &   No  &  Yes  &  Yes  &  $\fullcirc-\bigstar$    & 60.5   & 0.73 & 111 & -  & - \\
13 & $a4b2\phi180i150$  & Perturbation       &  No   &  No   &   No &   Yes &  $\fullcirc-\bigstar$  & 63.2    &  0.144  &   63.5 & $\Diamond$ & 192\\
14 & $a4b1\phi180i150$   & Disruption     &  $\bigstar$ &  Yes   &  No  &  Yes & $\fullcirc-\Diamond$ &   107 &   0.806 & 57.1 & - & - \\
15 & $a4b1/2\phi180i150$   & Disruption     &  $\bigstar,\Diamond$   &  Yes   &  Yes  &  No  &   - & -  & - & -  & $\fullcirc$ & 78.5 \\
16 & $a4b1/4\phi180i150$   & Exchange      &  No   &  No   &   No &  Yes   &  $\bigstar - \Diamond$  & 100   & 0.36 & 79.4 & $\fullcirc$  & 80 \\ 
\hline
17 & $a2b1/2\phi0i30$   & Disruption    &  $\Diamond$   &  No & Yes   & Yes &  $\fullcirc-\bigstar$    &  96.1  &  0.552  & 67.1 & - & - \\
18 & $a2b1/2\phi180i30$  & Merger   & No   &  Yes & Yes   & Yes &  $\fullcirc-(\bigstar + \Diamond)$    &  65.2  &  0.562  & 5.1 & - & - \\
19 & $a2b1/2\phi0i150$   & Disruption     &  $\bigstar,\Diamond$   &  No   &  Yes  &  No  &   - & -  & - & -  & $\fullcirc$ & 63.0 \\
20 & $a2b1/2\phi180i150$   & Disruption    &  $\bigstar,\Diamond$   &  Yes   &  Yes  &  No  &   - & -  & - & -  & $\fullcirc$ & 86.3 \\
21 & $a6b1/2\phi0i30$   & Disruption   &   $\Diamond$   &  No  &  Yes  &  Yes  &  $\fullcirc-\bigstar$   &  244 & 0.44 & 23 & - & - \\
22 & $a6b1/2\phi180i30$   & Merger     &  No   &   Yes  &  Yes  & Yes   &    $\fullcirc-(\bigstar + \Diamond)$   &  200  & 0.57  & 3.5 & - & -\\
23 & $a6b1/2\phi0i150$   & Perturbation  &  No   &   No  &  No  & Yes   &   $\fullcirc-\bigstar$ &  75.4  & 0.72  & 50 & $\Diamond$ & 150\\
24 & $a6b1/2\phi180i150$  & Disruption  & $\bigstar,\Diamond$    &   Yes  &   Yes &  No  &  -  & -   &  - & -  & $\fullcirc$ &  96.7 \\
\hline
25 & $a4b1/2\phi0i0$     & Disruption  &    $\Diamond$   &  No & Yes   & Yes &  $\fullcirc-\bigstar$    &  285  &  0.72  & 37.1 & - & - \\
26 & $a4b1/2\phi0i60$    & Disruption    &  $\Diamond$   &  No & Yes   & Yes &  $\fullcirc-\bigstar$    &  85  &  0.44  & 3 & - & - \\
27 & $a4b1/2\phi0i120^{\star}$    & Disruption (P)     &  $\Diamond$   &   No  & Yes   & Yes   &  $\fullcirc-\bigstar$  & - & - & - & - & - \\
28 & $a4b1/2\phi0i180$    & Disruption     &  $\Diamond$   &   No  & Yes   & Yes   &  $\fullcirc-\bigstar$  &  303  & 0.96 & 37 & -  & - \\ 

29 & $a4b1/2\phi180i0$   & Merger     &   No  &   Yes  &  Yes  &  Yes  &  $\fullcirc-(\bigstar + \Diamond)$   & 146  &  0.571  & 63 &  - & - \\
30 & $a4b1/2\phi180i60$   & Merger       &   No  & Yes    &  Yes  &  Yes  & $\fullcirc-(\bigstar + \Diamond)$     &   85.0 & 0.443 & 2.5 & - & - \\
31 & $a4b1/2\phi180i120$   & Merger (I)     & No    & Yes    &  Yes  &  Yes  &   $\fullcirc-(\bigstar + \Diamond)$   &  27.4  & 0.556 & 8 & - & -  \\
\hline
32 & $a4b1/2\phi45i30$    & Triple    &  No   &  No   &  No  &  Yes  &  ($\fullcirc-\Diamond$) - $\bigstar$  & 40.8  (91.5) & 0.44 (0.22) &  - & - & -\\
33 & $a4b1/2\phi90i30$     & Perturbation    &  No   &   No  &  No  &  Yes &  $\fullcirc-\bigstar$  &  38.0  &  0.272  & 76.2 &  $\Diamond$ & 180 \\
34 & $a4b1/2\phi135i30$   & Perturbation     &  No   &   No  &  No  &  Yes &  $\fullcirc-\Diamond$  &  43.6  &  0.80 & 62.1 &  $\bigstar$ &  186\\
35 & $a4b1/2\phi225i30$  & Triple    &  No   &  No   &  No  &  Yes  &   ($\bigstar-\Diamond)-\fullcirc$   &  58 (215)   & 0.49 (0.58) & - & - & -  \\
36 & $a4b1/2\phi270i30$   & Perturbation    &  No   &  No   &  No  &  Yes  &  $\fullcirc-\bigstar$  & 60.1   & 0.408 & 52 & $\Diamond$ & 153\\
37 & $a4b1/2\phi315i30$   & Merger     &   No  &  Yes   &  No  &  No & -  &  -   &  -   & - & $\fullcirc$, $\bigstar + \Diamond$ & 74, 71 \\
\hline
\end{tabular}
}
\caption{The outcomes of each simulated model: (\textit{first}) model number, (\textit{second}) outcome class, (\textit{third}) model name, (\textit{fourth}) whether the star survives or is destroyed, (\textit{fifth}) whether the two stars collide, (\textit{sixth}) whether the BH is accreting gas, (\textit{seventh}) whether a binary forms. A `(P)' next to the outcome class for Models 11 and 27 indicates partial disruption event and the designation `(I)' for Model 8 and 31 an interacting binary. The following four columns show the type of final binary product and its properties (semimajor axis, eccentricity, and velocity). The last two columns indicate the type of singles as a final product and their velocity. For triples, the semimajor axes and the eccentricities of the inner (outer) binaries are given without (within) a parenthesis. $\fullcirc$ indicates the $20\Msol$ BH initially in the binary, $\bigstar$ the $10\Msol$ star initially in the binary, and $\Diamond$ the incoming single star. The two models 11 and 27 with their names having a superscript $^{\star}$ are the cases where the final outcomes are not determined until the end of the simulations because of two prolonged interactions. In those two models, an unstable triple forms with the outer object on a very eccentric orbit so that strong interactions are expected as soon as the outer object returns to the inner binary. For this case, we only provide the type of the inner binary members. }\label{tab:outcome}
\end{table*}

\section{Methods}\label{sec:method}

Our numerical methods are essentially the same as described in \citetalias{Ryu+2023b}, except that the incoming object is now a $10\Msol$ main-sequence (MS) star. We concisely summarize the key elements here, but we refer to \citetalias{Ryu+2023b} for the specific details. We perform a suite of 3D hydrodynamic simulations of the close encounters using the massively parallel gravity and magnetohydrodynamic moving mesh code {\small AREPO} \citep{Arepo,ArepoHydro,Arepo2}. We use the {\small HELMHOLTZ} equation of state \citep{HelmholtzEOS} which includes radiation pressure, assuming local thermodynamic equilibrium. 

The initial state of the two stars, the one in the binary and the incoming single, is identical and was taken from evolved MS stars with the core H mass fraction of 0.3 (at age of 18 Myr) computed using the stellar evolution code {\sc MESA} (version r22.05.1) \citep{Paxton+2011,paxton:13,paxton:15,MESArelaxation,paxton:19,jermyn22}. This stellar model is the same as the one adopted in \citetalias{Ryu+2023b}. We refer to the Section 3.2 `Stellar model' in \citetalias{Ryu+2023b} for the choices of the parameters adopted to evolve the star and  for its radial density profile. We map the 1D {\sc MESA} model into a 3D {\small AREPO} grid with $N\simeq 5\times10^{5}$ cells, and fully relax the resulting 3D single star.

We model the BH using an initially non-rotating sink particle, which interacts gravitationally with the gas and can grow in mass via gas accretion. We follow exactly the same procedure for accretion described in \citetalias{Ryu+2023b}, including the refinement criteria introduced there. 

We parameterize the binary's semimajor axis $a$ using the analytical approximation of the Roche lobe radius by \citet{Eggleton1983}, 
\begin{align}
   \frac{r_{\rm RL}}{a}= \frac{0.49 q^{2/3}}{0.6q^{2/3}+\ln(1+q^{1/3})},
\end{align}
where $r_{\rm RL}$ is {the volume averaged Roche lobe radius of the star}, $q=M_{\star}/M_{\bullet}$ is the mass ratio, and $a$ is the orbital separation. We define $a_{\rm RL} \equiv a(R_{\rm RL}=R_{\star})$ as the separation at which the star fills its Roche lobe. For $q=0.5$ and $r_{\rm RL} = R_{\star}$,  $a_{\rm RL} \simeq 3.12$, and $R_{\star}\simeq 16.9\Rsol$. 

\subsection{Initial conditions}\label{subsec:initial}

In our simulations, a circular binary consisting of a $20\Msol$ BH and a $10\Msol$ star encounters a single $10\Msol$ star on a parabolic trajectory. As remarked in \citetalias{Ryu+2023b}, the choices of the encounter parameters are somewhat arbitrary, but BHs with such masses have been observed in X-ray binaries \citep[e.g.][]{binder_wolfrayet_2021}. In addition, encounters between objects of similar masses are expected in the centers of young star clusters where massive objects accumulate due to mass segregation. Later, based on our simulation results, we discuss potential effects of different masses in \S~\ref{subsec:differentmass}.

We consider three semi-major axes: $a/a_{\rm RL}=2$, $4$ and $6$, corresponding to orbital periods of 4, 12, and 22 days, respectively. The distance between the binary's center of mass and the BH at the first closest approach is parameterized using the impact parameter $b$, i.e., $r_{\rm p}=ab/2$. Here, $r_{\rm p}$ is the pericenter distance and $a$ the binary semimajor axis. 

We investigate the dependence of encounter outcomes on key encounter parameters, i.e., inclination angle $i=0$, $30^{\circ}$, $60^{\circ}$, $120^{\circ}$ and $180^{\circ}$, $b = 1/4$, 1/2, 1 and 2, and the phase angle $\phi=0^{\circ}- 315^{\circ}$ with an increment of $\Delta\phi=45^{\circ}$. Here, $\phi$ is the initial angle between the line connecting the two members in the binary and the $x-$axis (see Figure 2 in \citetalias{Ryu+2023b}). To study the impact of $\phi$, we initially rotate the binary while the initial separation between the binary and the star is fixed at $5a$. We study the dependence of outcomes on $i$, $\phi$ and $b$ using the encounters of the intermediate-size binaries ($a/a_{\rm RL}=4$).

We summarize the initial parameters considered in our simulations in Table~\ref{tab:initialparameter}. Each of the models is integrated to a few up to $100\,t_{\mathrm p}$, which is the typical time it takes to identify the final outcomes. Here, $t_{\rm p}= (r_{\rm p}^3/GM)^{1/2}$ is the dynamical time at $r=r_{\rm p}$ and $M$ is the total mass of three objects. The value of $t_{\rm p}$ for each model is given in Table~\ref{tab:initialparameter}.

\section{Results}\label{sec:results}

\subsection{Classification of outcomes}\label{sub:class}

We divide the outcomes produced in the three-body encounters between BH-star binaries and single stars into five classes: \textit{Stellar disruption}, \textit{Merger}, \textit{Orbital perturbation}, \textit{Member exchange}, and \textit{Triple formation}. We provide a sketch with an overview of the five outcome classes with corresponding short descriptions in Figure~\ref{fig:summary}, and we summarize the outcome types and properties for all our models in Table ~\ref{tab:outcome}. 

\begin{enumerate}
    \item \textit{Stellar disruption}: this class refers to encounters in which one or both stars are fully destroyed via tidal disruptions and collisions with the BH. These disruptive encounters mostly take place when the BH and the single star meet first. Once the star is destroyed, the BH is quickly surrounded by an accretion flow, which would generate an electromagnetic transient (see Section~\ref{subsec:accretion}). Disruptions can also happen when two stars encounter first on a retrograde orbit ($i=150^{\circ}$): in two models (Models 15. $a4b1/2\phi180i150$ and 20. $a2b1/2\phi180i150$), the two stars merge first, followed by the disruption of the merged star by the BH.  
    The three types of final outcomes in this class are:
    \begin{itemize}
        \item Single BH: a single accreting BH when both stars are destroyed (Models. 15. $a4b1/4\phi180i150$, 19. $a2b1/2\phi0i150$, and 20. $a2b1/2\phi180i150$). The ejection velocity of single BHs is $60-100\,{\rm km\, s^{-1}}$, which is high enough to escape globular clusters. 
        
        \item Binary: a full disruption of the incoming star and the original binary where the BH is accreting, hence there is no ejected single star (e.g., Models 3. $a4b1/2\phi9i30$ and 4. $a4b1/4\phi0i30$). 
        
        \item Binary + single: a partial disruption event of the incoming star, creating an unbound partially disrupted star, and the original binary where the BH is accreting (e.g., Models 11. $a4b1/2\phi0i150$ and 27. $a4b1/2\phi0i120$).
        
    \end{itemize}

    \item \textit{Merger}: this corresponds to the case where the two stars merge and survive. These events frequently occur when the two stars encounter first on a prograde orbit ($i=30^{\circ}$). For this case, the merged stars form a binary with the BH, except in one case where the merged star and the BH are unbound (Model 36. $a4b1/2\phi315i30$). The semimajor axes and the eccentricities of the binaries are $a\simeq 130 - 200\Rsol$ and $e\simeq0.5-0.7$, respectively. 

    \item \textit{ Orbit perturbation}: in this class, the incoming star weakly perturbs the orbit of the original star and becomes an unbound single. The final outcome is the perturbed binary consisting of its original members and an ejected star which was the incoming single. We do not identify a well-defined region of the parameter space which is specific to this class: the encounter parameters of these cases cover almost the entire range considered in this work. The perturbed binaries have a smaller $a$ than the initial value by $10 - 40$ percent, depending on the encounter parameters. The ejected stars have a velocity ranging between $120 - 230\,{\rm km\,s^{-1}}$. 

    \item \textit{ Member exchange}: We find that in five models (out of 37) the initial binary is dissociated and a new binary forms while the third object is ejected. The newly formed binaries consist of either the BH and the initially incoming star (Models 5. $a4b2\phi180i30$, 8. $a4b1/4\phi180i30$, 34. $a4b1/2\phi135i30$) or the two stars (Models 2. $a4b1\phi0i30$ and 16. $a4b1/4\phi180i30$). The semi-major axes of the newly formed binaries are larger than that of the original binary by $5-50$ percent in all these models except Model 34. $a4b1/2\phi135i30$ where the newly formed binary is smaller than the original binary by 30 percent. The eccentricities of the newly formed binaries are in the range $0.1-0.8$. The ejection velocities of the singles vary between $80 - 230\,{\rm  km\, s^{-1}}$.  Given our small sample size, we could not reliably identify the parameter region where the member exchange happens frequently. 

    \item \textit{Triple formation}: In this class, a hierarchical triple forms after the original binary is dissociated (Models 1. $a4b2\phi0i30$, 32. $a4b1/2\phi45i30$, and 35. $a4b1/2\phi225i30$). In two cases (Models 1. and 35.), the inner binary consists of the two stars and the tertiary is the BH. In the last model, the star originally in the binary is in an outer orbit around the inner binary made up of the BH and the initially single star.  According to the stability criteria by \citet{Vynatheya+2022}, which is an improved version of \citet{MardlingAarseth2001}, all these triples are unstable. 
    
\end{enumerate}

\begin{figure*}
	\centering
	\includegraphics[width=8.6cm]{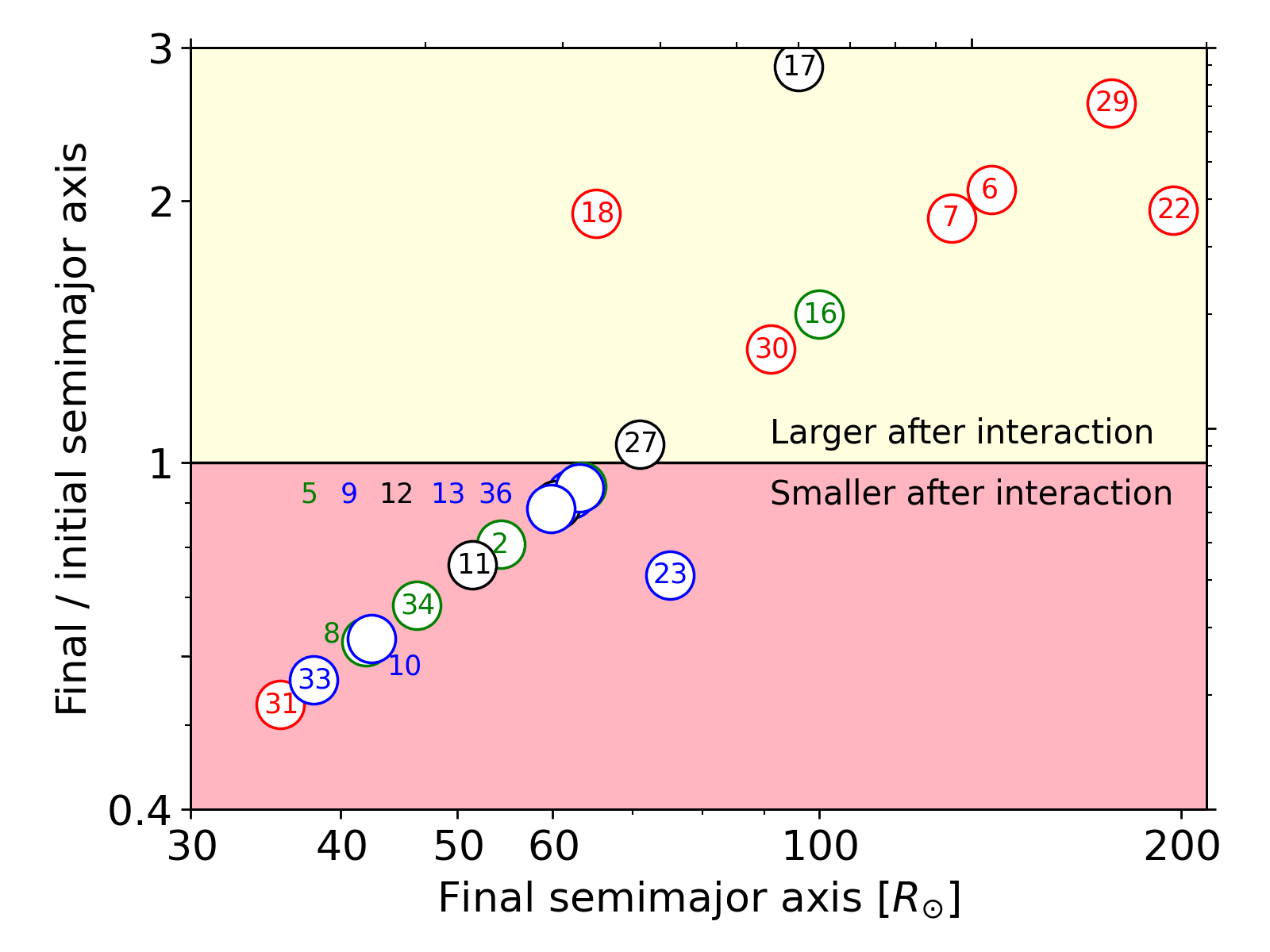}
	\includegraphics[width=8.6cm]{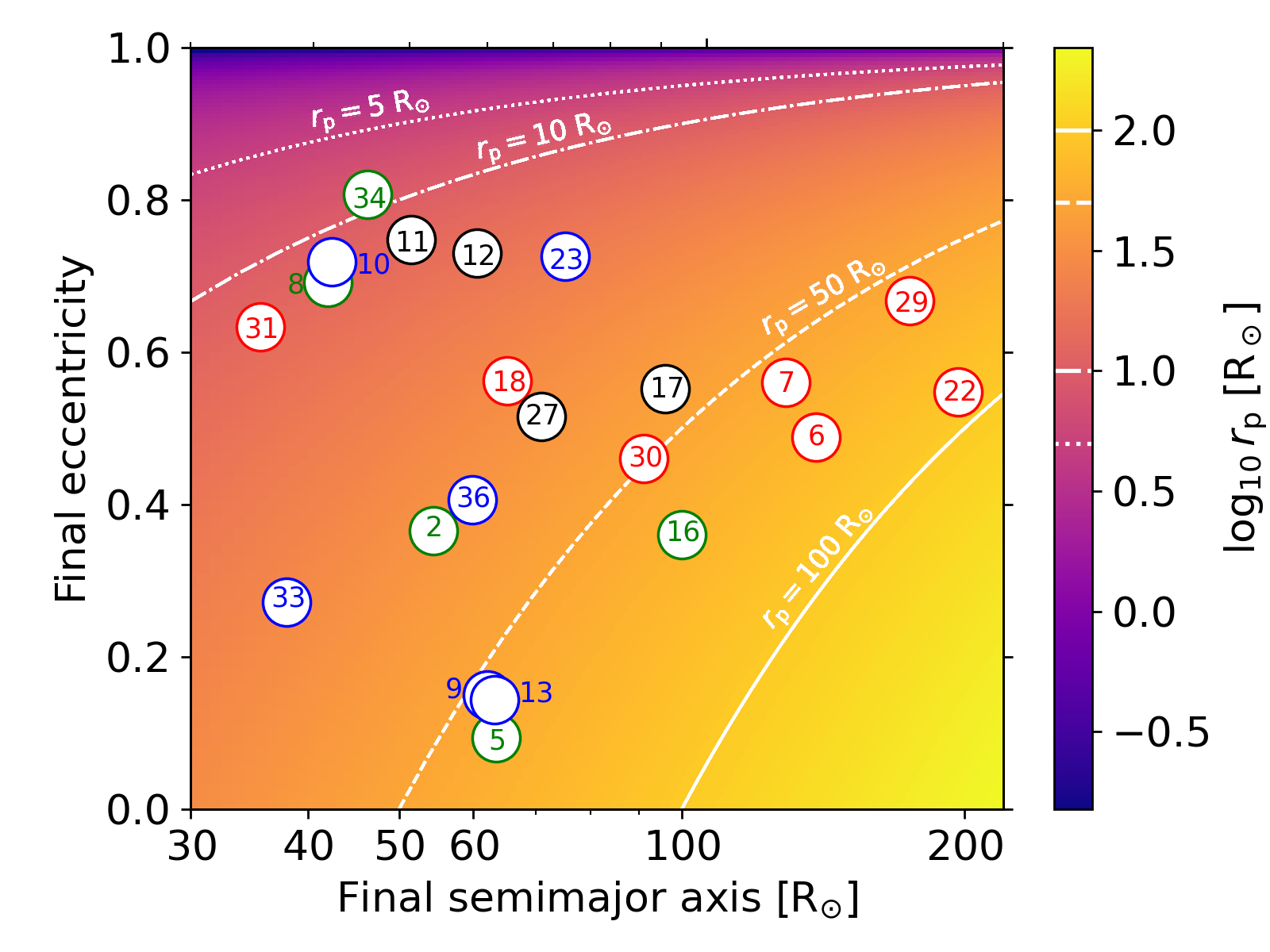}
\caption{Semimajor axis $a$ and eccentricity $e$ of the final binaries: (\textit{left}) the ratio of the final $a$ to the initial $a$ as a function of the final $a$, and (\textit{right}) final $e$ as a function of the final $a$. The number in each circle indicates the model number, whereas its color shows the category in which the outcomes of the corresponding model fall into: Black (\textit{disruption}), red (\textit{merger}), blue (\textit{orbit perturbation}), and green (\textit{member exchange}). The plot in the \textit{right} panel shows the pericenter distance $r_{\rm p}$ using a logarithmic scale with four guiding lines, indicating $r_{\rm p}=5.4\Rsol$ (dotted, the stellar radius of the original star), $10\Rsol$ (dot-dashed), $50\Rsol$ (dashed), and $100\Rsol$ (solid). Note that the power-law relation with small scatter in the \textit{left} panel is simply because the initial semimajor axis is the same in most models (or $a = 4\,a_{\rm RL}$). The four binaries with $a > 200 \Rsol$ are excluded in this plot to focus on the parameter space occupied by the majority of the final binaries. }
	\label{fig:binaryproperty}
\end{figure*}

\begin{figure*}
	\centering
	\includegraphics[width=18cm]{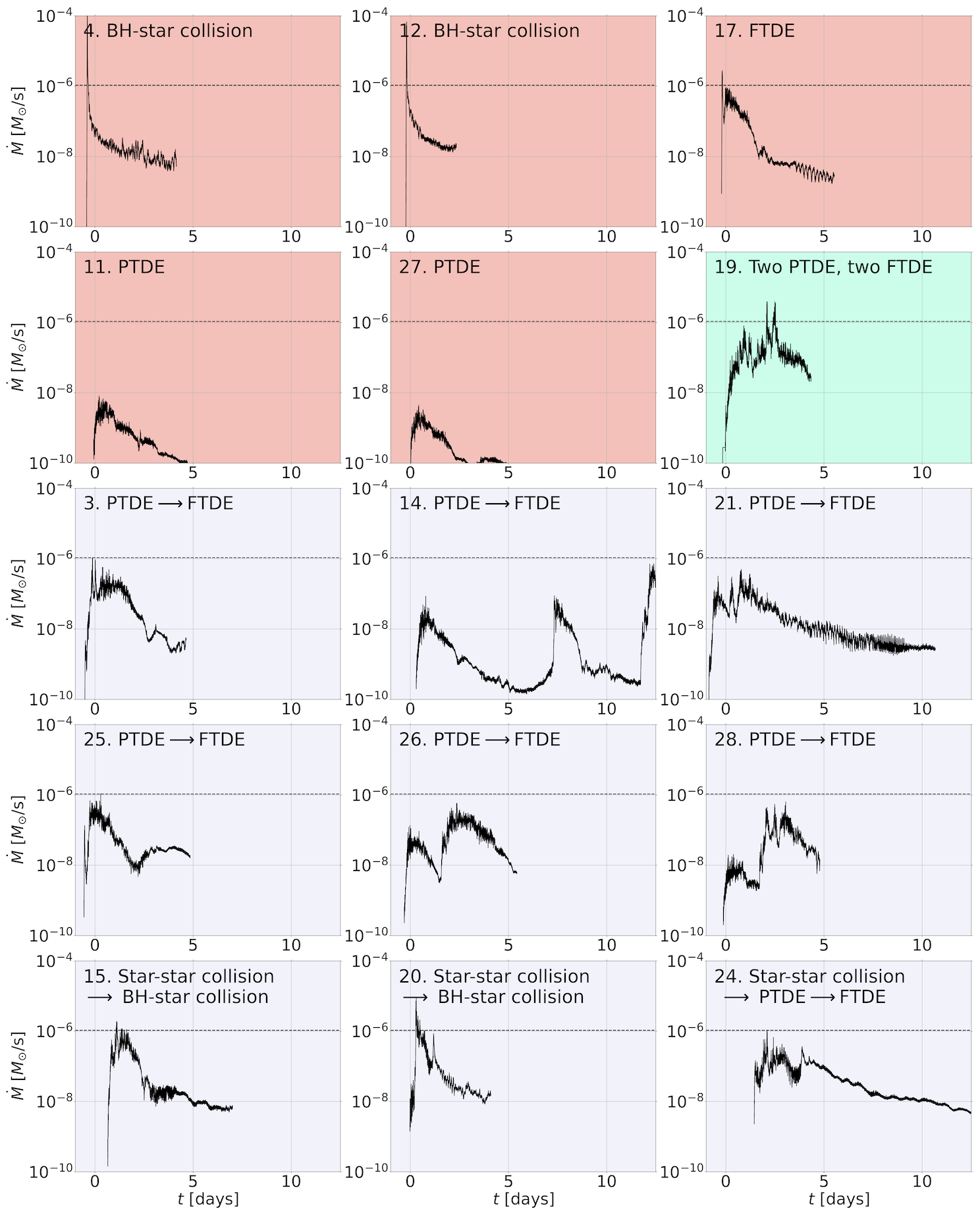}
\caption{Accretion rate of the black hole in 15 models where an accretion disk forms via various combinations of stellar disruptions and mergers. The mechanisms that drive the accretion processes are written next to the model numbers. Here, FTDE (PTDE) refers to full (partial) tidal disruption events. We classify the rate curves depending on the number of disruptive events and the resulting shape using different background colors, as follows: single event (red), multiple events leading to relatively flat rates (green), multiple events leading to multiple peaks (blue). The black dashed line in each panel indicates the median of the peak accretion rates of all the models.  }
	\label{fig:accretionrate}
\end{figure*}

\begin{figure}
	\centering
	\includegraphics[width=8.6cm]{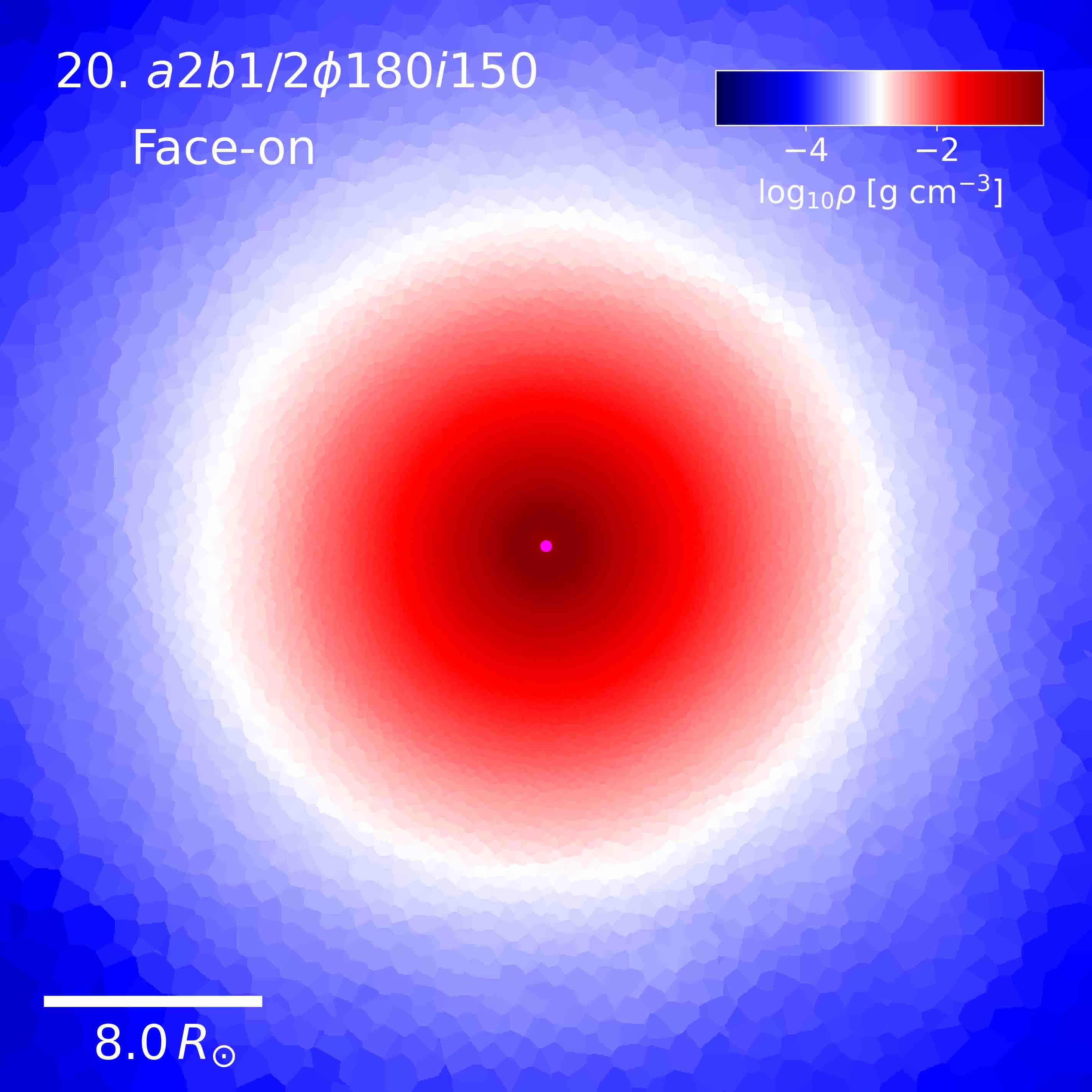}
    \includegraphics[width=8.6cm]{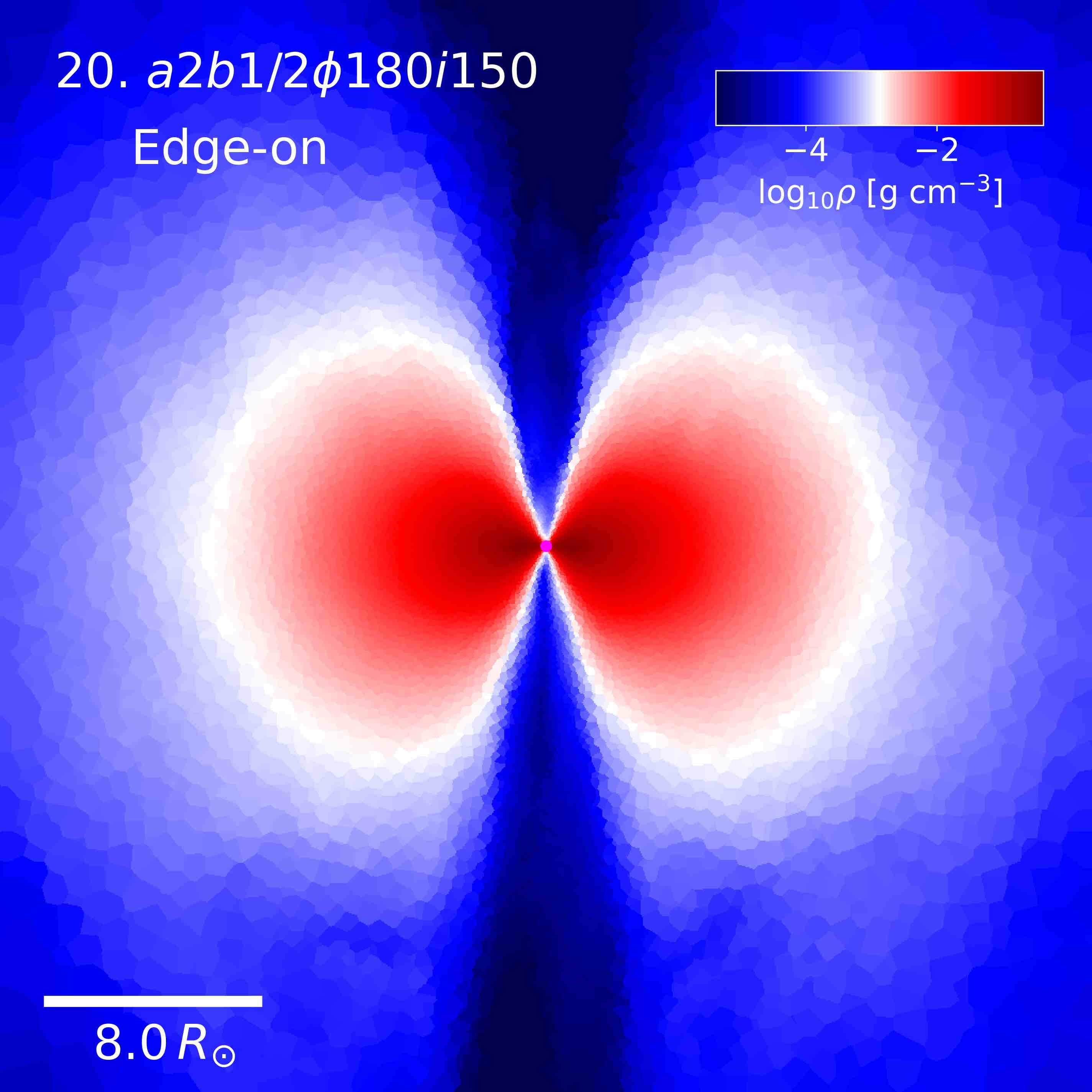}
\caption{Density distribution of the disk in Model 20. $a2b1/2\phi180i50$ in the equatorial plane (\textit{top}) and in a vertical plane (\textit{bottom}). }
	\label{fig:diskdensity}
\end{figure}

\begin{figure}
	\centering
	\includegraphics[width=8.6cm]{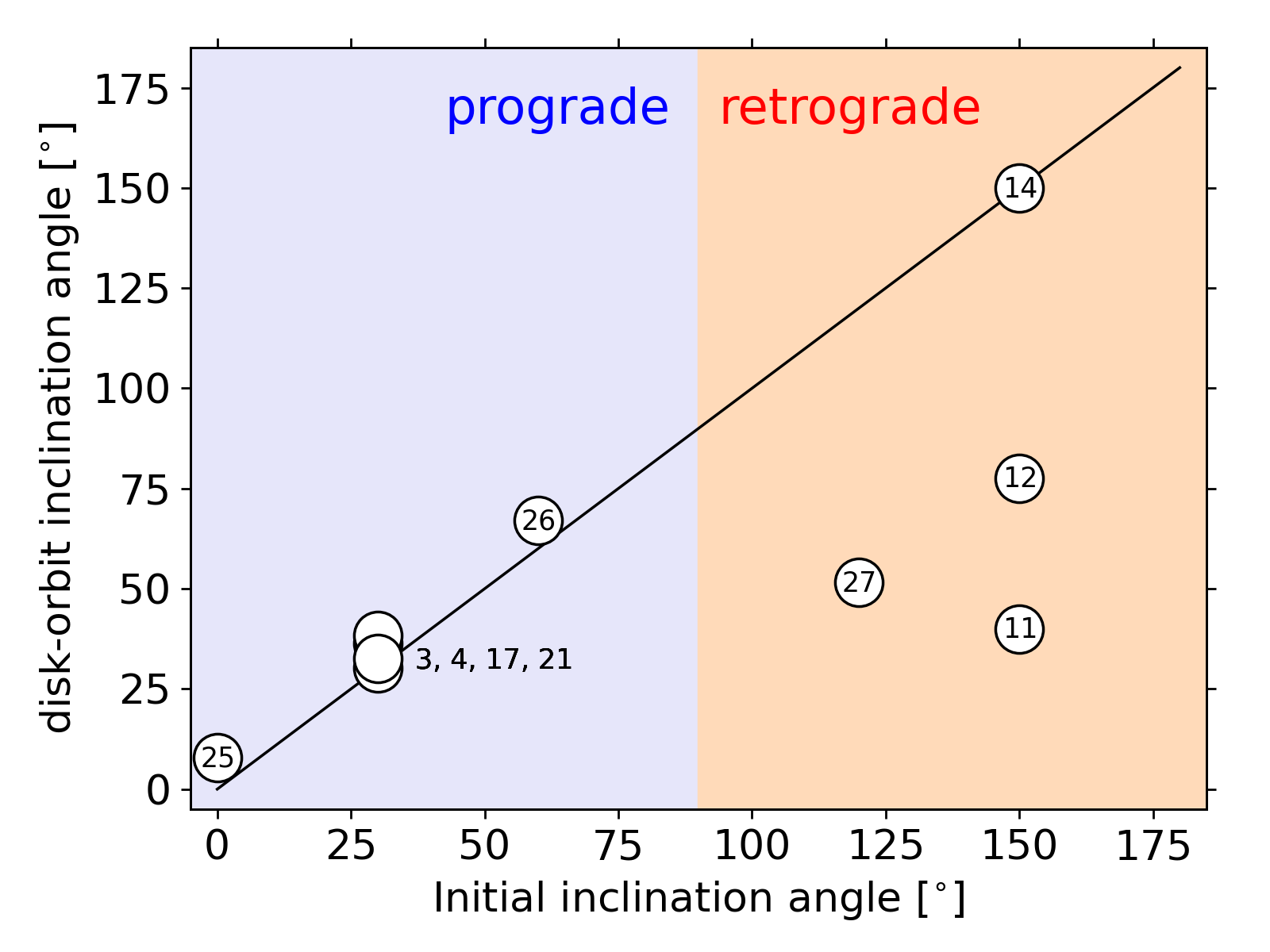}
\caption{The mutual inclination angle between the accretion disk spin axis and the BH-star orbital axis, as a function of the initial encounter inclination, in models where the final outcome is a binary with the BH surrounded by a disk. We excluded Model 28 because the star in the final binary would be partially destroyed at the next pericenter passage. }
	\label{fig:inclination}
\end{figure}

\subsection{Dependence of outcomes on parameters}\label{subsec:dependence}

\begin{enumerate}
    \item \textit{Phase angle}: This primarily determines which two objects meet first. As indicated by the varieties of the outcomes (e.g., tidal disruption, triple, binary, and merger) from the models with varying $\phi$ (Models 3 and 32-37), the outcomes sensitively depend on the exact configuration at the first encounter. A general trend is that the chances of having mergers are significantly higher in encounters where the two stars meet first, compared to the cases where the BH and the incoming star interact closely first. For the latter, a likely outcome is stellar disruption. For the parameters covered by Models 3 and 32-37, a full disruption occurs within a relatively small region of $\phi$ ($\Delta \phi < 45^{\circ}$). 

    \item \textit{Impact parameter}: Violent events (i.e., tidal disruption , collision, and merger) tend to occur when the impact parameter is less than $a/2$ (or $b <1$). However, a small impact parameter does not always lead to such star-destroying events, depending on other encounter parameters, primarily the phase angle. Hence $b <1$ is a necessary condition for disruptive interactions. For example, in Model 16. $a4b1/4\phi180i150$, the incoming star dissociates the binary, but the interactions do not ultimately lead to either a stellar disruption or merger. 

    \item \textit{Inclination angle}: Two primary effects of the inclination angle are as follows. Firstly, the inclination angle determines how small the relative velocity becomes between the two meeting objects, which is directly translated into the size of the gravitational-focusing encounter cross section: the higher the relative velocity (retrograde), the smaller the cross section is. For the same encounter parameters, prograde encounters likely create star-removing events. Second, although rare, a high relative speed in retrograde encounters inversely indicates that, if a strong encounter occurs, the resulting change in the momentum would be relatively high. In fact, because the head-on collision of the two stars so effectively removes the kinetic energy of the two colliding stars, the merger events followed by a disruption only occurs in retrograde cases.  

    \item \textit{Semimajor axis}: Encounters involving an initially smaller binary (e.g., $a/a_{\rm RL}=2$) appear to more preferentially create TDEs and collisions, which may be attributed to the fact that the incoming star would be more directed towards the binary's center of the mass where violent interactions are more likely to occur. However, we do not see any other clear trend associated with the size of the binary. 
\end{enumerate}

\subsection{Binary formation}\label{subsec:binary}

In our simulations, the formation of a binary, as a final product, is very frequent. We present in Figure~\ref{fig:binaryproperty} the orbital properties of the binaries. As shown in the \textit{left} panel, this type of three-body interactions results in both wider and more compact binaries than the initial binaries with $a\lesssim 100\Rsol$. The semi-major axes of most of the final binaries range from $35\Rsol$ to $200\Rsol$. Our simulations also show that wide binaries with $a$ as large as $a\simeq 1000\Rsol$ can be produced by three-body interactions involving a relatively compact binary. However, we could not find any significant dependence of the ratio of the final $a$ to the initial $a$ on any of the parameters or the types of outcomes introduced in Section~\ref{sub:class}. Note that the well-defined power-law relation with small scatter is simply because of the same initial semi-major axis in most models (or $a = 4\,a_{\rm RL}$). 

The final eccentricities, as shown in the \textit{right} panel of Figure~\ref{fig:binaryproperty}, are within $e\simeq 0.1 - 0.8$ for binaries with $a\lesssim 200\Rsol$. The wide binaries with $a\gtrsim 300 \Rsol$ are more eccentric, $e\simeq 0.86 - 0.99$. Note that the star in the very wide binary in Model 28. $a4b1/2\phi0i180$ will be partially disrupted at the next pericenter passage because the pericenter distance ($\simeq 4\Rsol$) is between its full disruption radius $\simeq 3\Rsol$ and the partial disruption radius $\simeq 14\Rsol$\footnote{The full disruption radius is calculated using Equation 5 in \citet{Ryu+2020a} and the partial disruption radius using Equation 17 in \citet{Ryu+2020b} assuming the star has the same initial structure as the original star. }. Similarly to the semi-major axis, we do not find any clear trends of the final eccentricity in terms of the encounter parameters and final outcomes. This may imply that the properties of the final binaries are sensitively dependent on multiple encounter parameters. We also find that interacting binaries form in Models 8. $a4b1/4\phi180i30$ and 31. $a4b1/2\phi180i120$, the models with (I) next to the class in Table~\ref{tab:outcome}.

\begin{figure*}
	\centering
	\includegraphics[width=8.6cm]{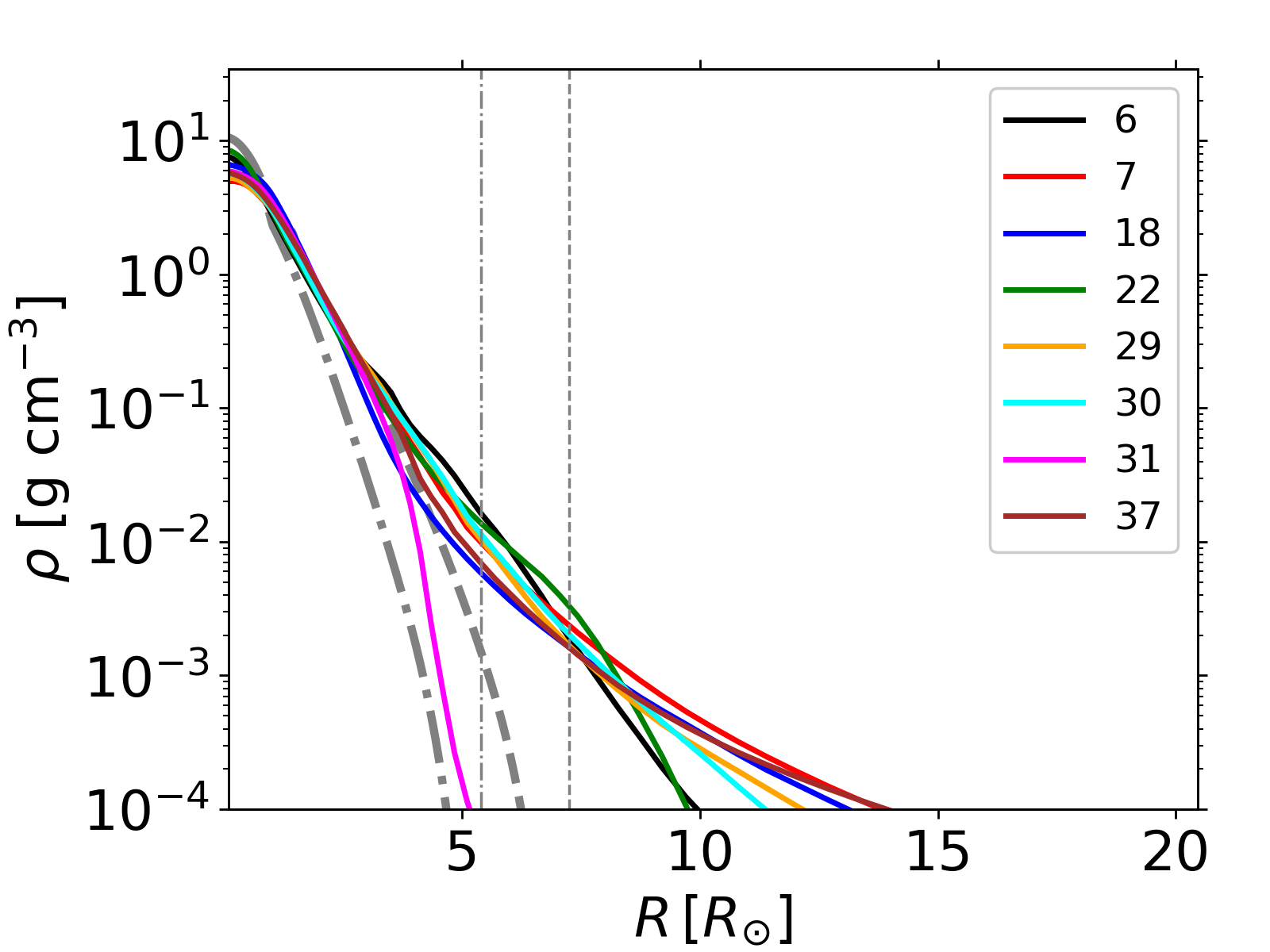}
	\includegraphics[width=8.6cm]{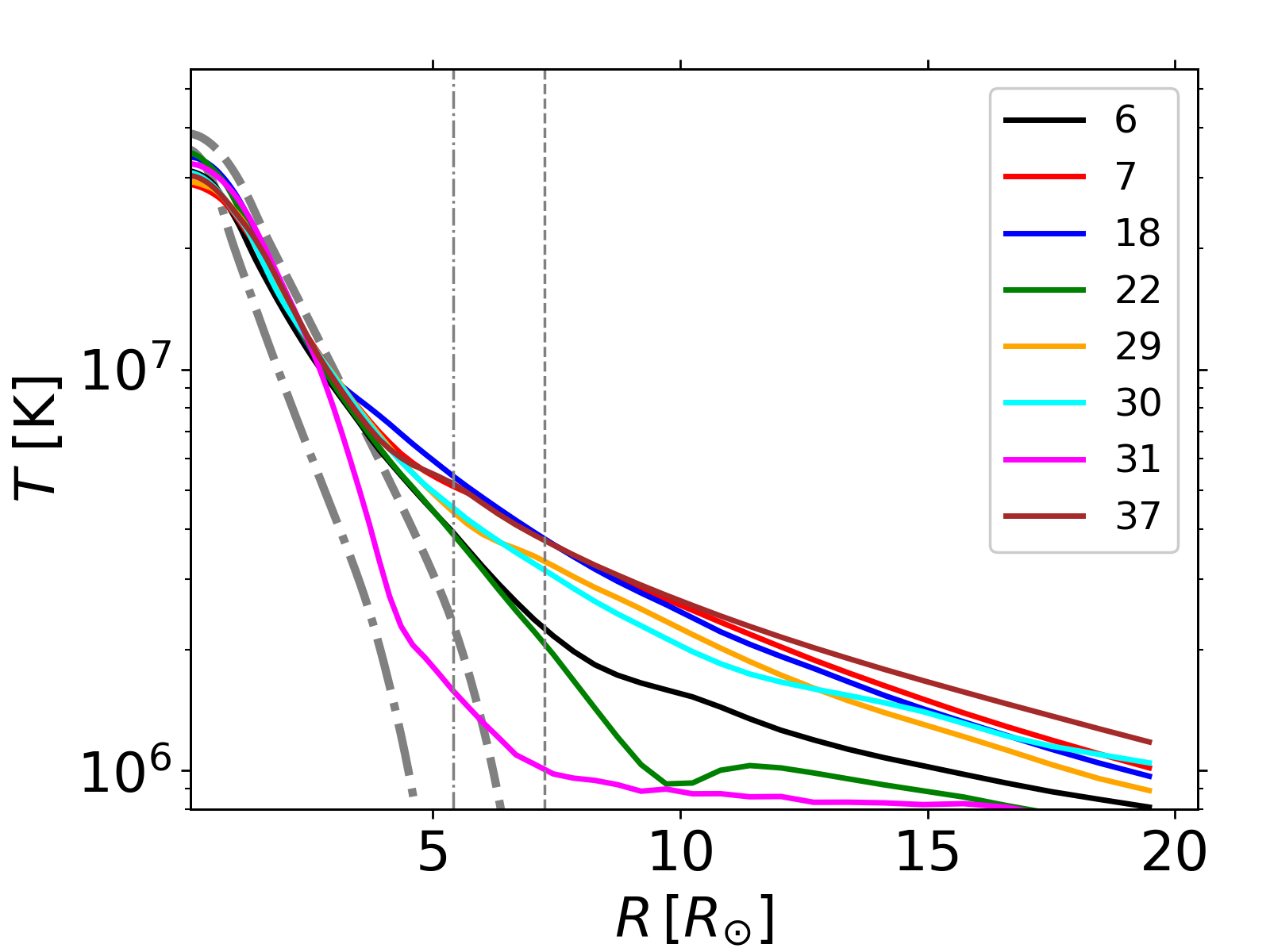}\\
 	\includegraphics[width=8.6cm]{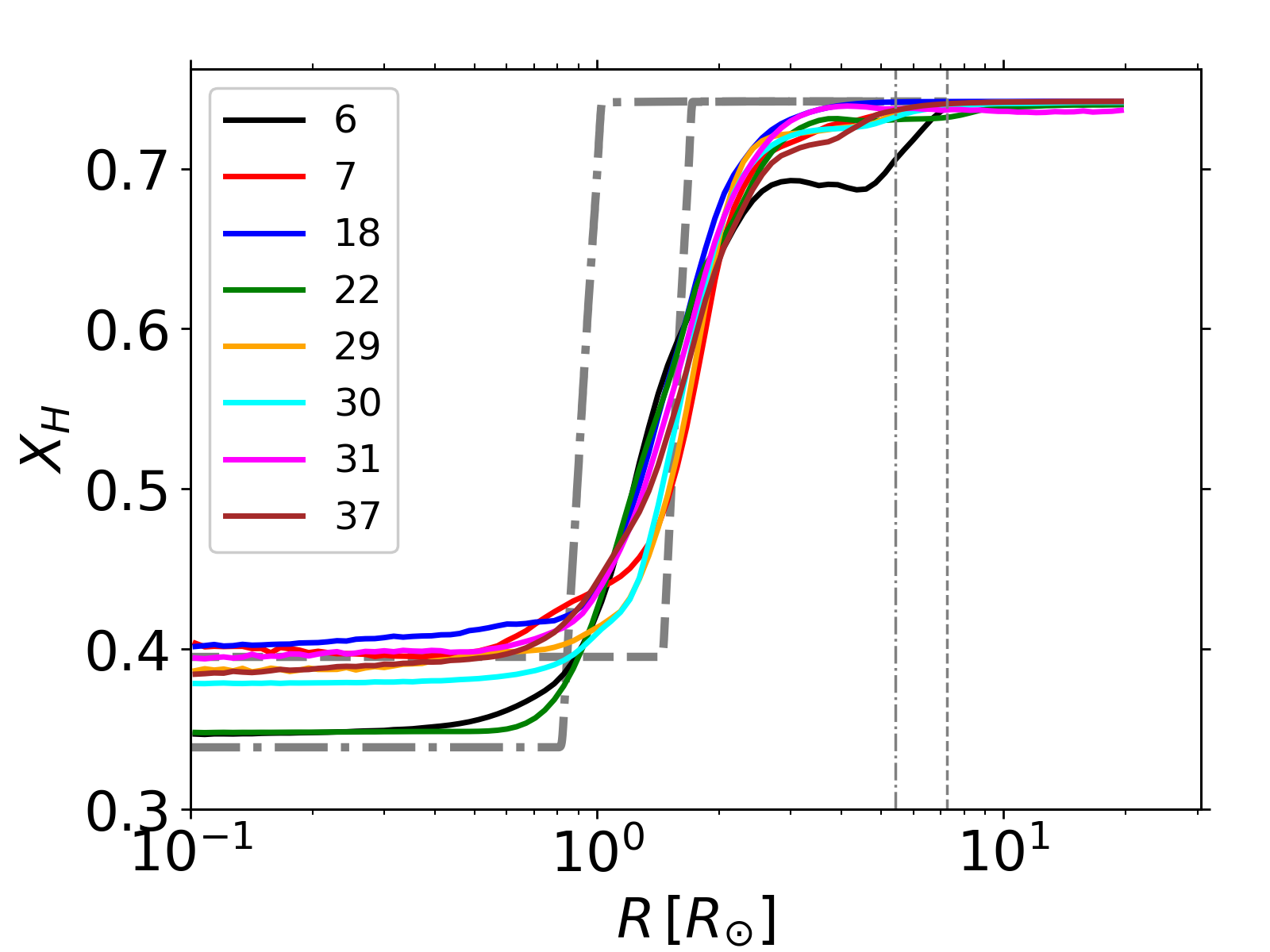}
	\includegraphics[width=8.6cm]{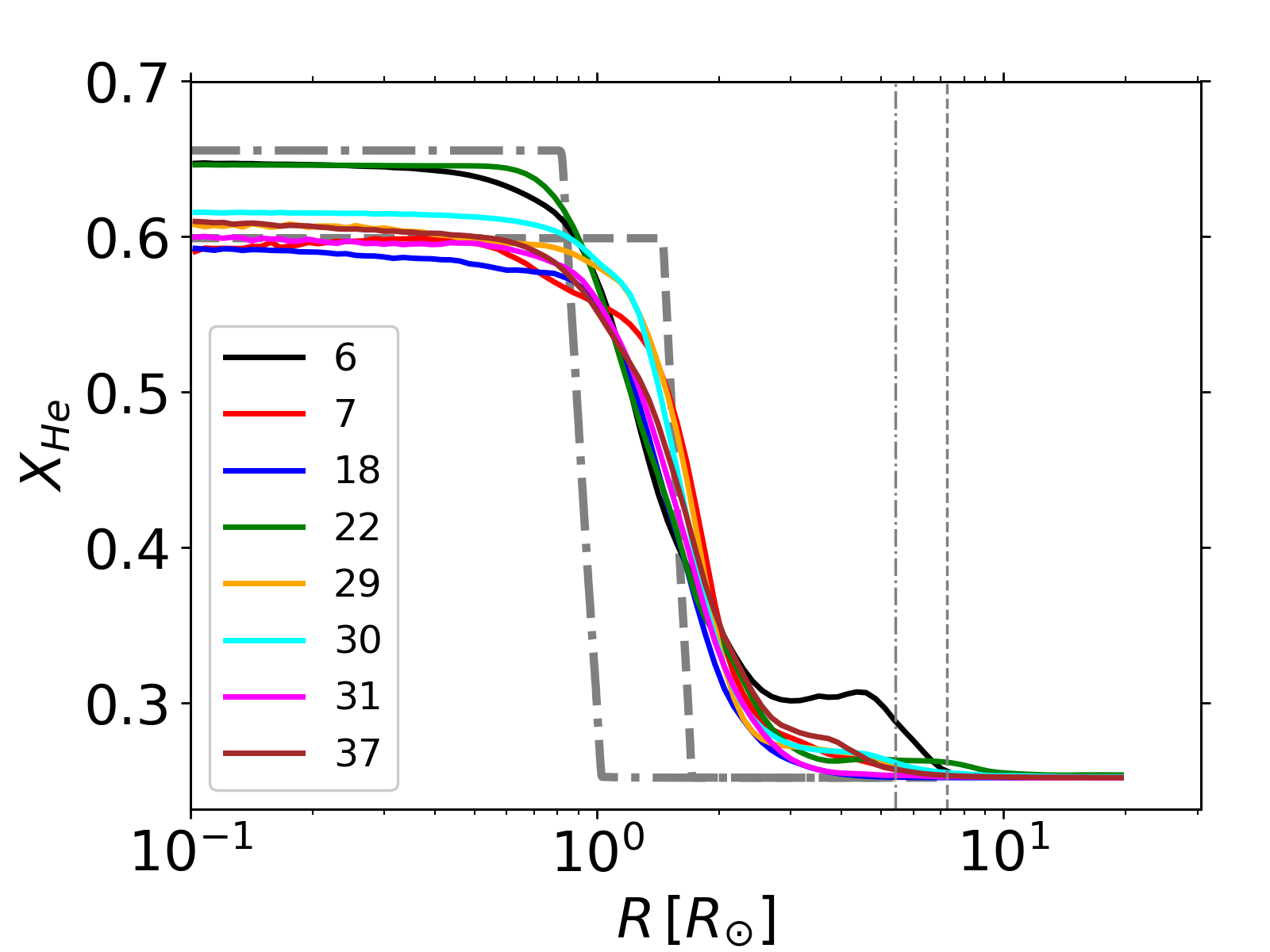}\\
\caption{Density (\textit{top-left}), temperature (\textit{top-right}), and chemical component distribution (\textit{bottom-left}: H and \textit{bottom-right}: He) of merged stars as a function of radial distance from the core. The two grey thicker lines in each panel show, respectively, the profile of the initial $10\Msol$ star (dot-dashed) and of ordinary non-rotating MS stars with a mass of $18.5\Msol$ (dashed), which roughly corresponds to the average mass of the merged stars. The vertical grey thinner lines, matching the line styles, indicate the radii of those ordinary stars: $5.4\Rsol$ for the $10\Msol$ star (dashed), and $7.3\Rsol$ for the $18.5\Msol$ star (dot-dashed).}
	\label{fig:mergerproduct}
\end{figure*}

\begin{figure*}
	\centering
\includegraphics[width=4.4cm]{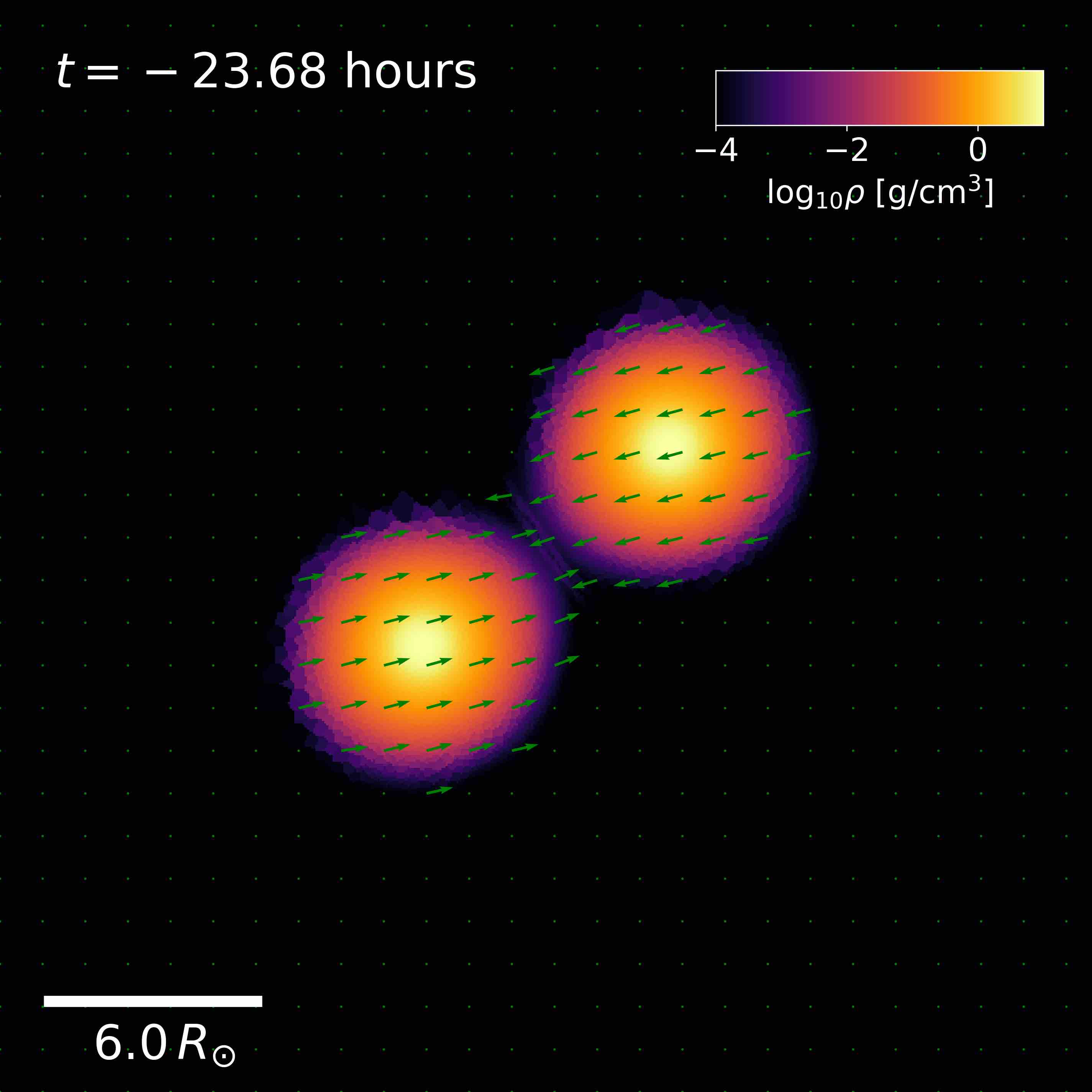}
\includegraphics[width=4.4cm]{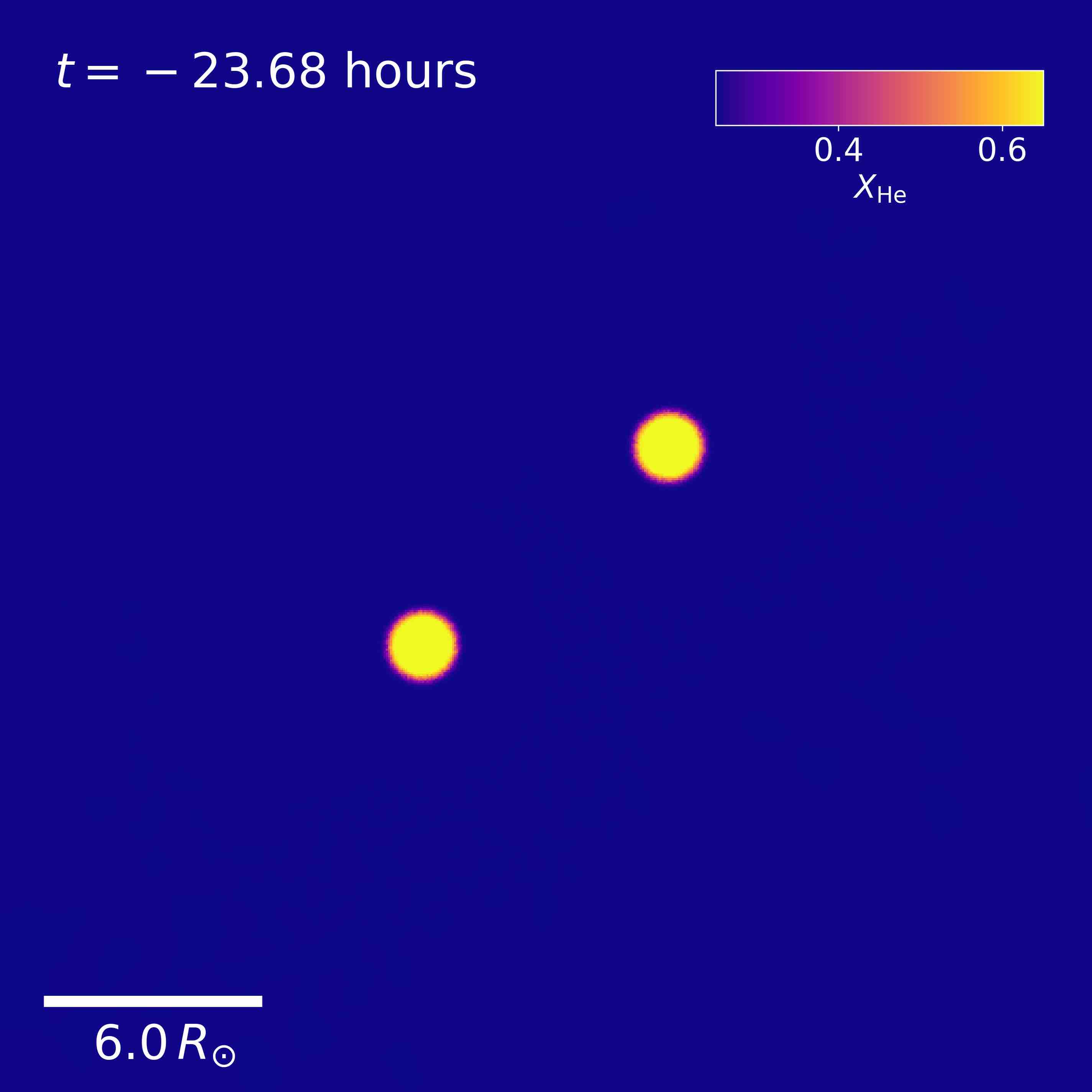}\\
\includegraphics[width=4.4cm]{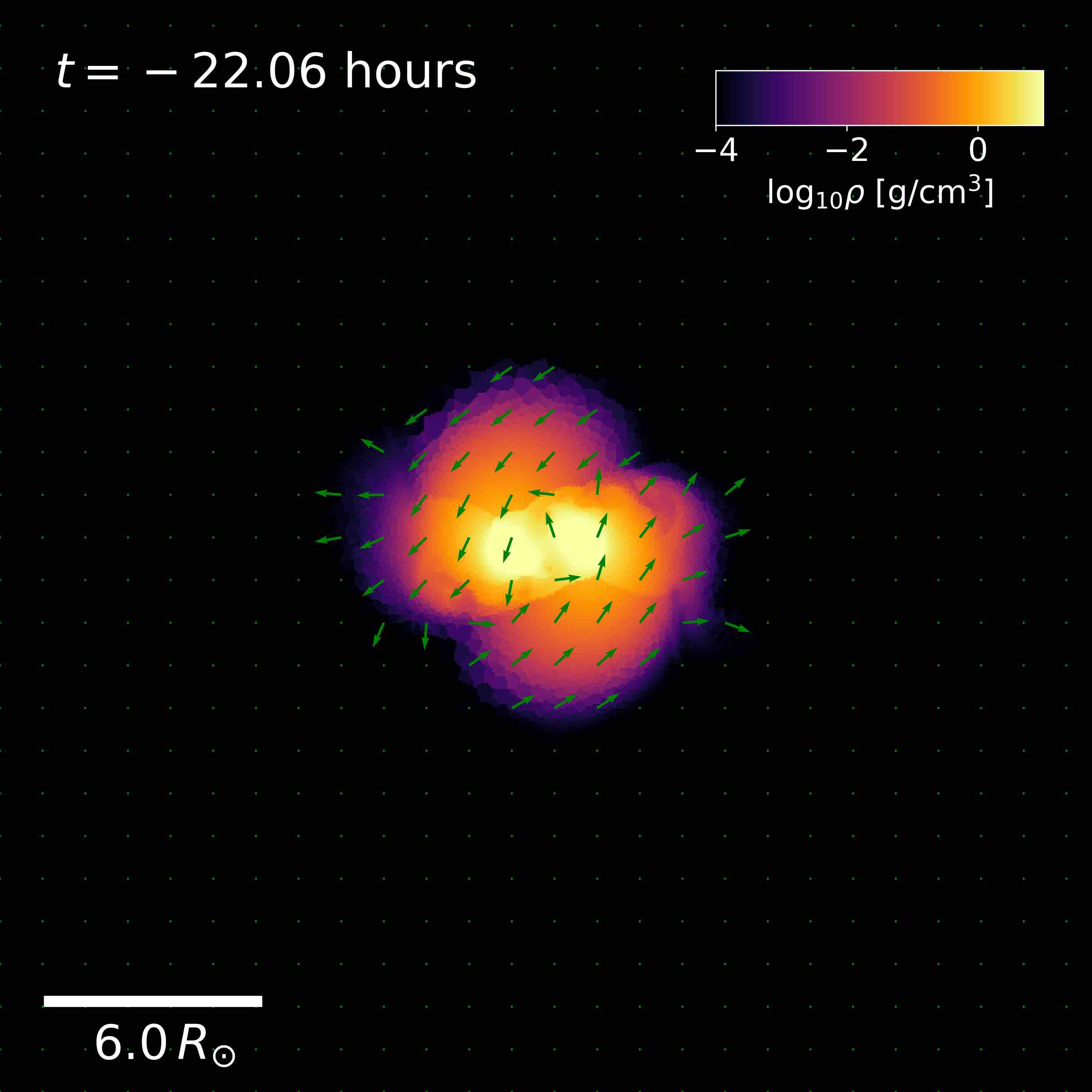}
\includegraphics[width=4.4cm]{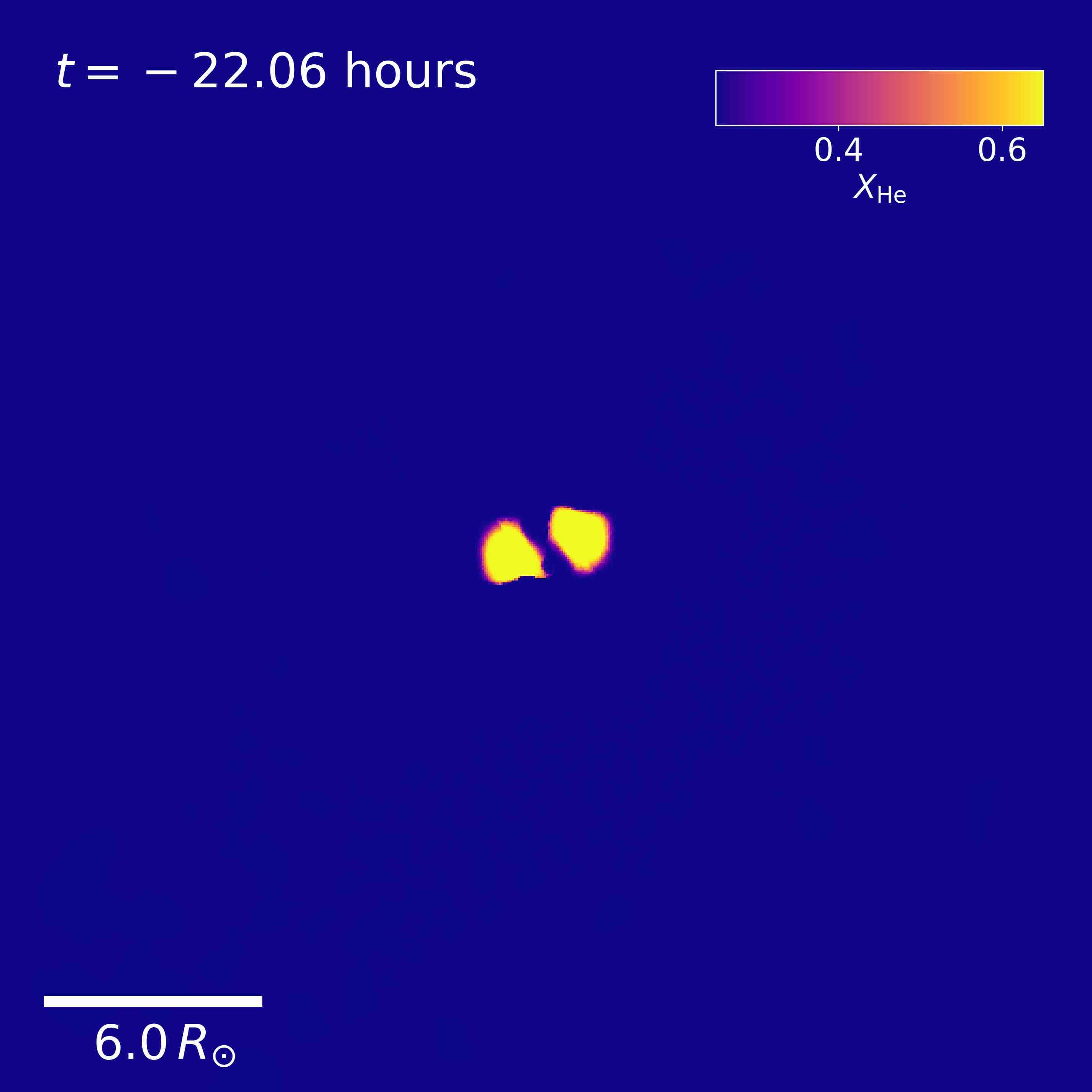}\\
\includegraphics[width=4.4cm]{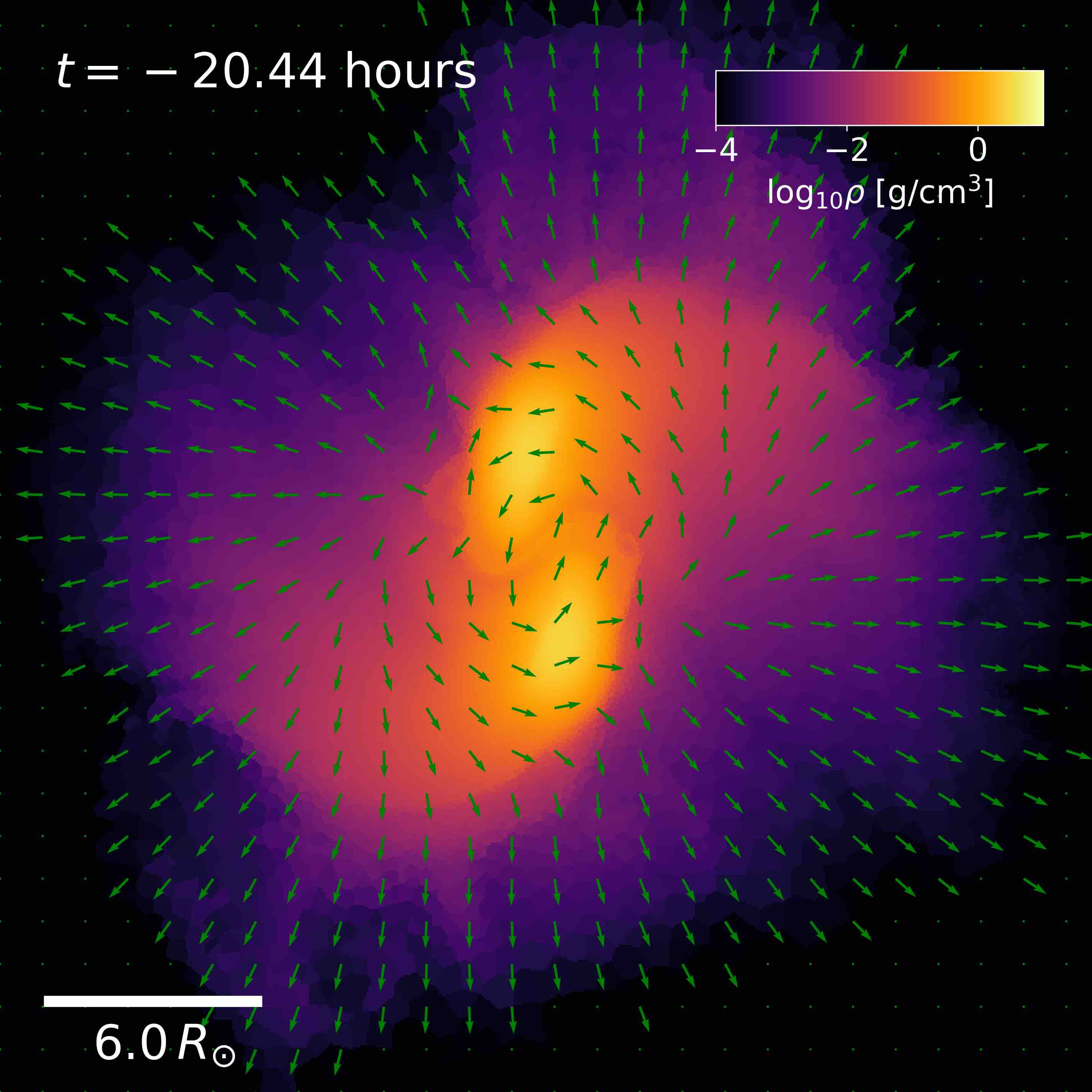}
\includegraphics[width=4.4cm]{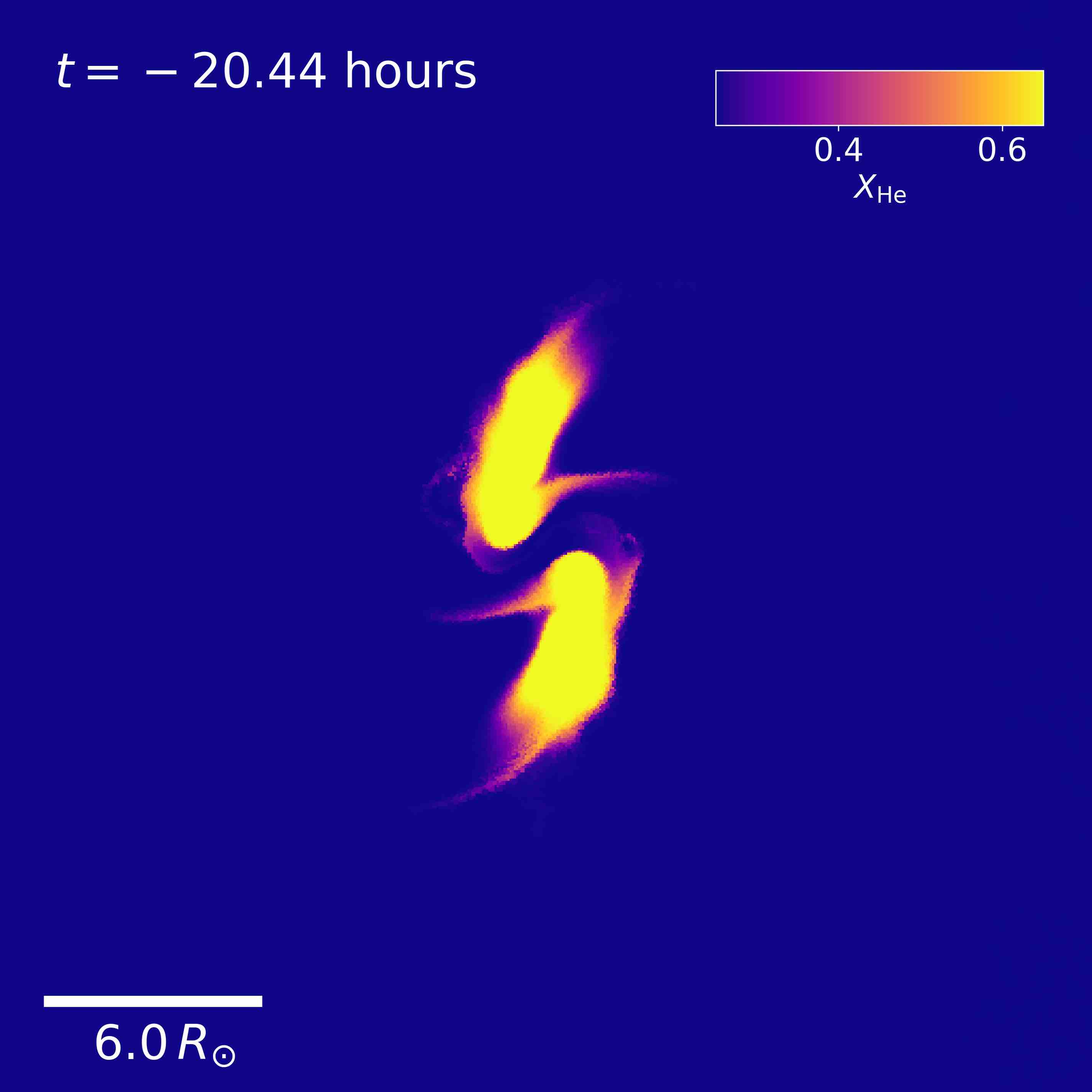}\\
\includegraphics[width=4.4cm]{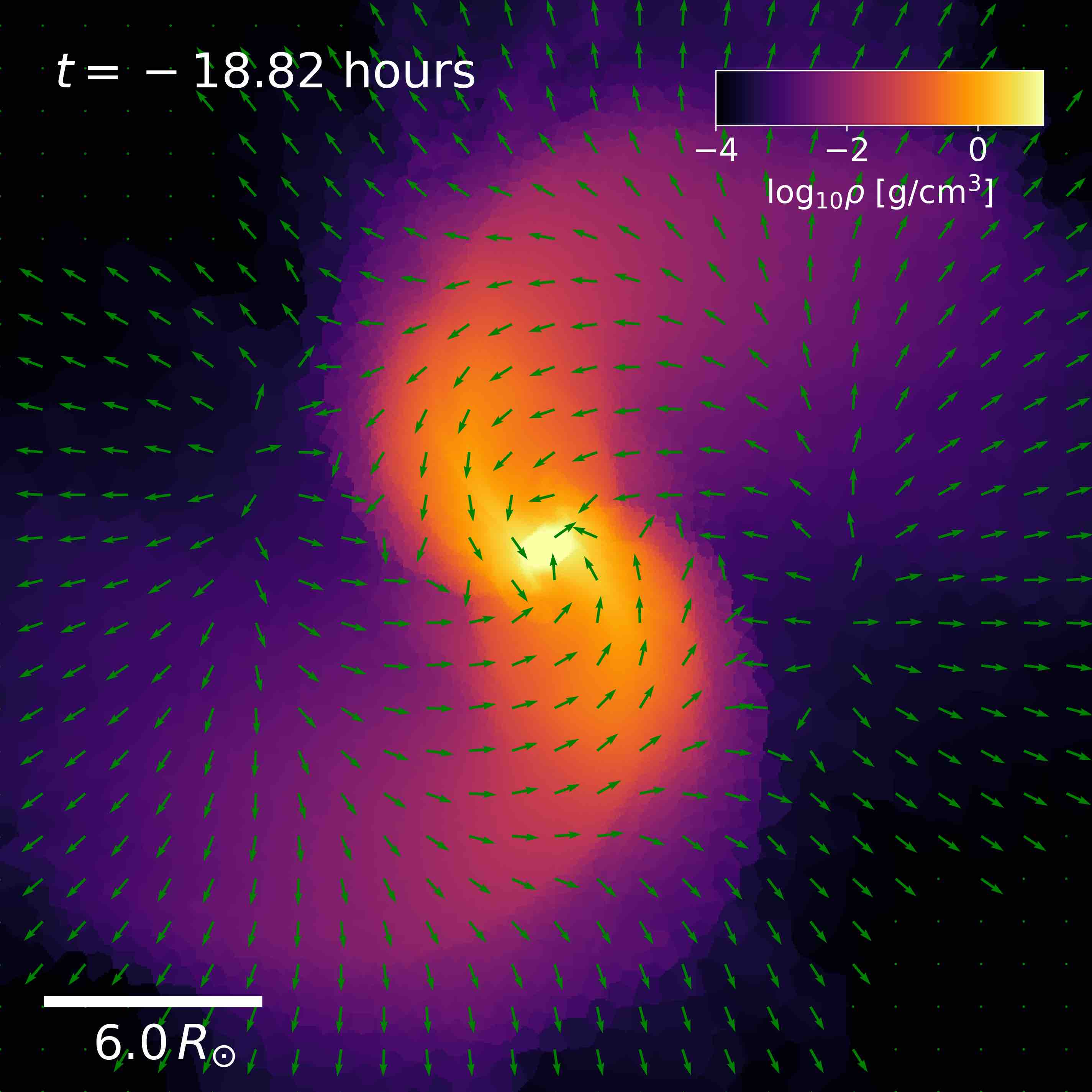}	
\includegraphics[width=4.4cm]{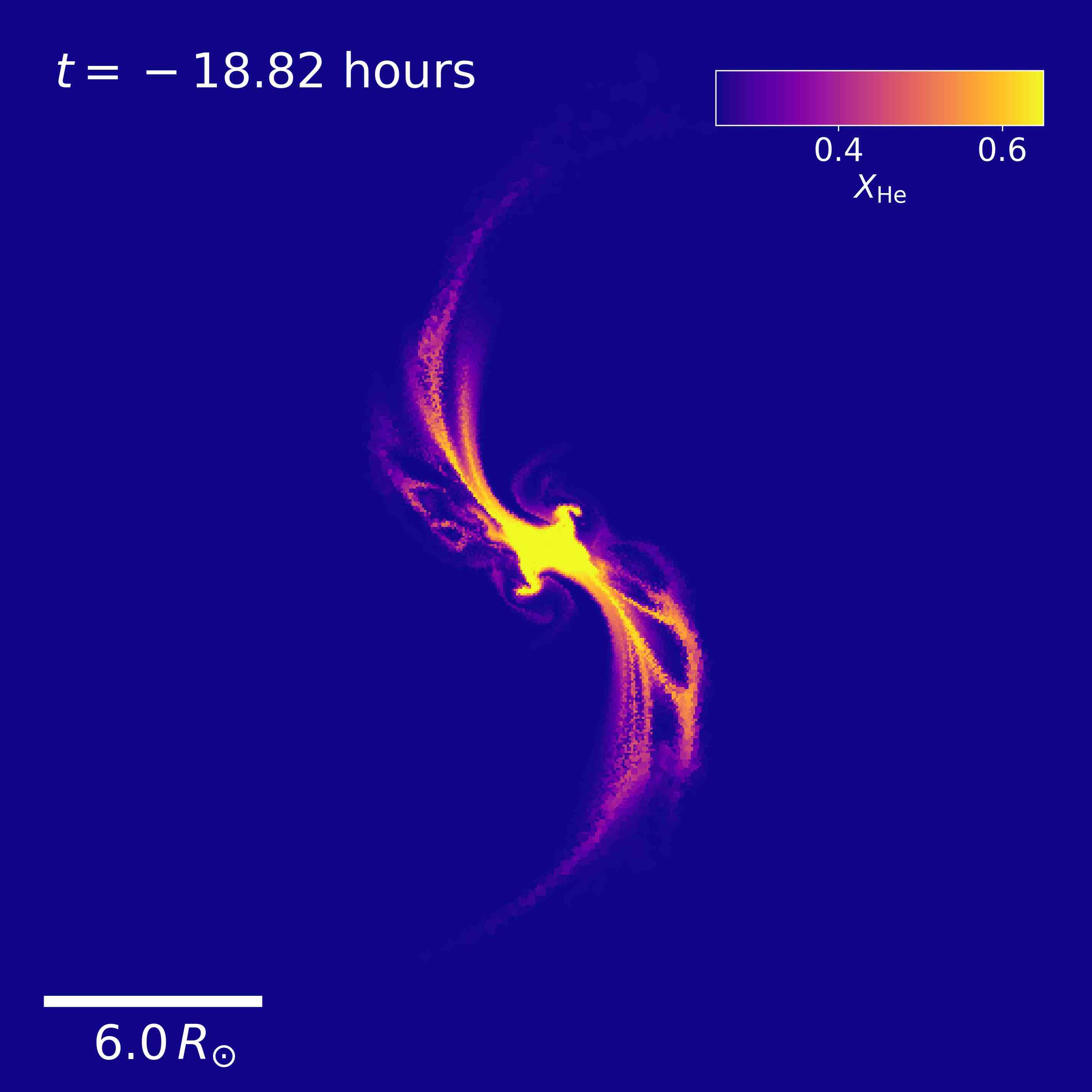}	\\
\includegraphics[width=4.4cm]{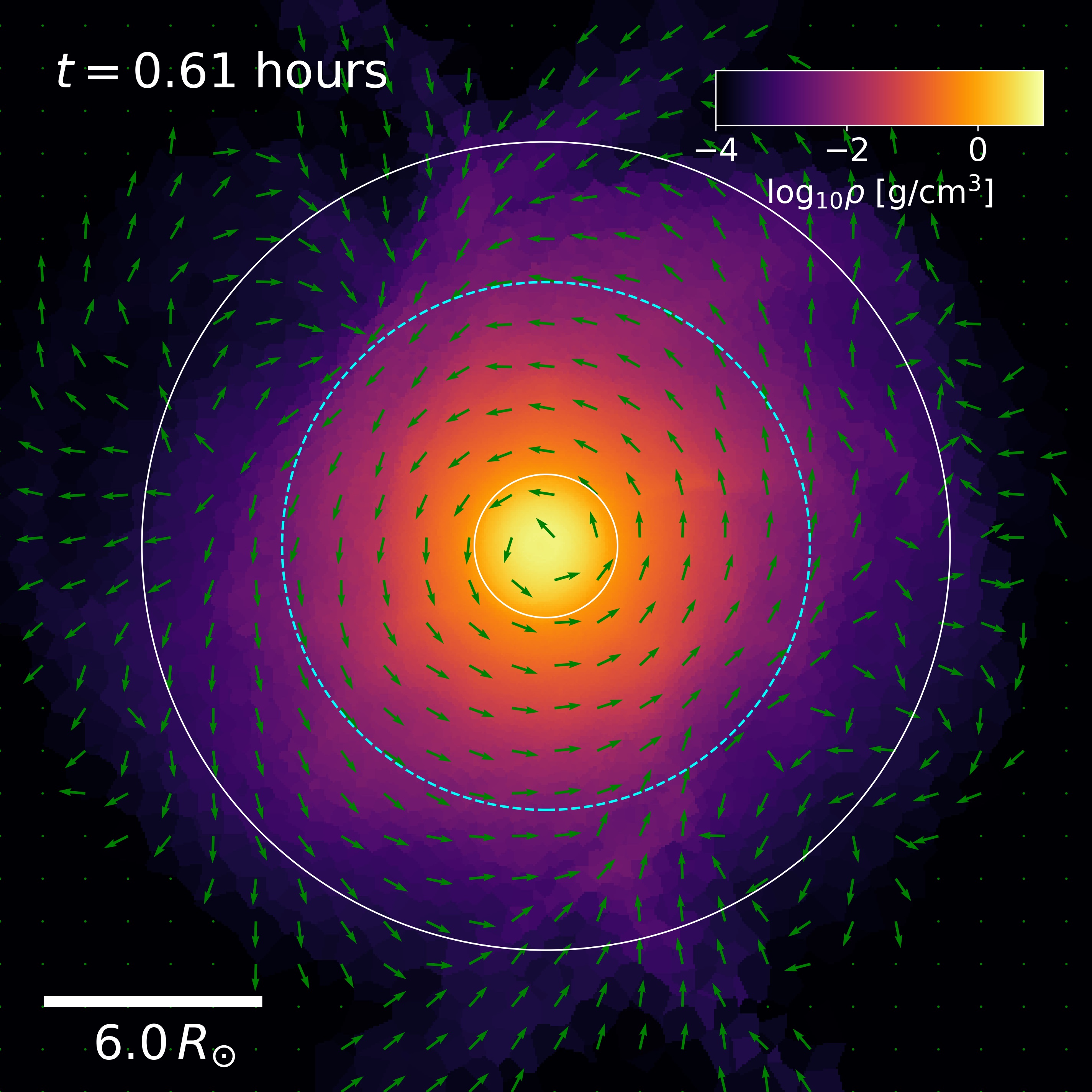}
\includegraphics[width=4.4cm]{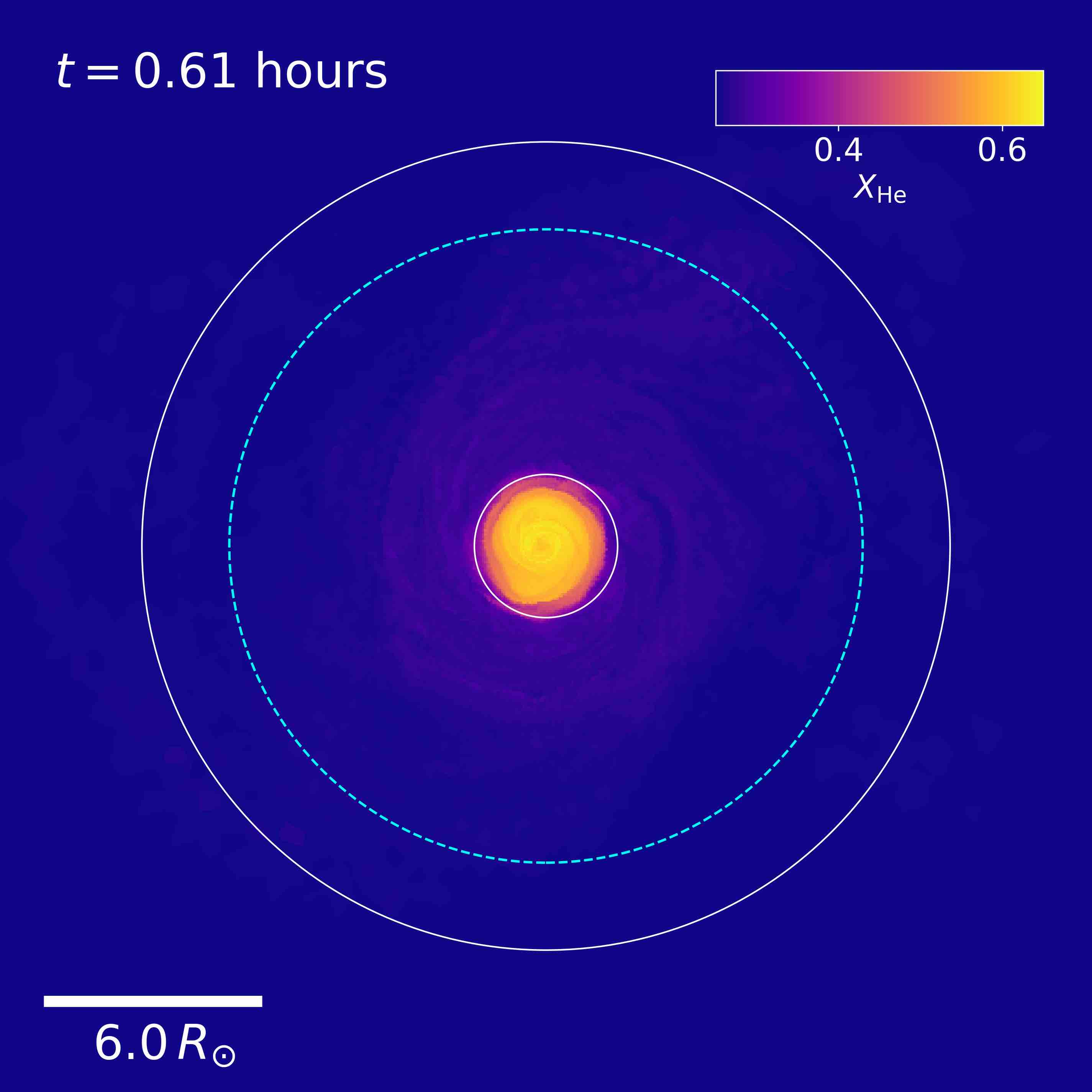}
	\caption{Density (\textit{left}) and Helium abundance (\textit{right}) distribution in the equatorial plane in an off-axis nearly parabolic collision in Model 29. $a4b1/2\phi180i0$ at five different times in the center of mass frame of the two colliding stars. The arrows in the left panels indicate the motion of gas.}
	\label{fig:mergerproduct3}
\end{figure*}

\subsection{Accretion}\label{subsec:accretion}

In the majority of our models, the BHs are surrounded by gas produced in TDEs, collisions, and stellar mergers. In those cases, the BH accretes gas, potentially creating electromagnetic transients (EMTs), although there may be a significant delay between the moment of the close encounter and the  peak emission of the EMT because of a large optical depth of the debris (or a long cooling time). We present in Figure~\ref{fig:accretionrate} the accretion rate $\dot{M}$ in models where at least one of the stars is disrupted. As shown in the figure, the shape of the accretion rate as well as the peak rate, ranging from $10^{-9}-10^{-4}\Msol\,{\rm s}^{-1}$, are diverse. We  split the types of $\dot{M}$ curve into three categories, depending on their shape and the mechanism that creates the accretion disk.
\begin{itemize}
    \item Single-peak :  $\dot{M}$ rises relatively rapidly and decays slowly, which can be generated in two cases. 1) Partial TDE (Models 11 and 27): in this event, only a fraction of mass is lost from a star (most often the incoming one), which quickly forms an accretion disk. The peak accretion rate is substantially lower than in other cases, $\dot{M}\lesssim 10^{-7}\Msol\,{\rm s}^{-1}$. 2) Full TDE or head-on collision (Models 4, 12, and 17): when the incoming star undergoes a collision with the BH or is completely tidally disrupted, the accretion rate surges very rapidly and then decays.  

    \item Multiple-peaks: To produce an $\dot{M}$ with more than one peak, more than one violent event should occur. We find three such cases in our simulations. 1) Partial TDE $\longrightarrow$ full TDE (Models 3, 14, 21, 25, 26, and 28): in this case, a partial TDE occurs, followed by a full disruption. 2) Merger$\longrightarrow$ full TDE (Models 15, 20, and 24): when the two stars merge, some fraction of mass is ejected (see~Section\ref{subsec:merger}). The BH nearby captures the gas and accretes it (e.g., the first $\dot{M}$ peak at $t\simeq 1$ days in Model 20). If the collision significantly reduces the kinetic energy of the merged star, this places the merged star on a radial orbit around the BH and it is disrupted at the first pericenter passage. For this case, the time difference between peaks is determined by how far from the BH a merger happens and how quickly the merged star is disrupted.

    \item Rise-flat: $\dot{M}$ in Model 19 rises on a time scale of $1$ day and then stays nearly constant at $\dot{M}\simeq 10^{-7}\Msol\,{\rm s}^{-1}$. The flat $\dot{M}$ indicates that gas is continuously injected into the BH. In fact, in this model the two stars are partially destroyed at each pericenter passage, soon followed by two total disruptions. As a result of continuous mass inflow into the BH, the overall shape of the accretion rate is flat. 
\end{itemize}

In the remainder of this section, we investigate the properties of the accretion disk around the BH. The accretion disk is sub-Keplerian and optically and geometrically thick. As an example, we depict the density of the disk formed in Model 20. $a2b1/2\phi180i150$ in Figure~\ref{fig:diskdensity}. In general, the disks have an aspect ratio of $0.4 - 0.6$ at distance $r\gtrsim 1\Rsol$ from the BH, which increases inwards to $\gtrsim 1$ at $r\lesssim 0.1\Rsol$. The azimuthal velocity of the disks is $\simeq 0.6 -0.9$, indicating that the disk is radiation pressure-supported. The density of the disks is mostly flat at $r\lesssim 0.1-1\Rsol$, and it decreases outwards following a power-law of $r^{-3}-r^{-4}$. The temperature decreases approximately monotonically as $r$ increases: $T\propto r^{-0.25}$ at $r < 1\Rsol$ and $T\propto r^{-1}$ at $r> 1\Rsol$. The $r-$ scaling relations for $\rho$ and $T$ are very similar to those for the disks that form in three-body interactions between BH-star binaries and single BHs (see Figure 9 in \citetalias{Ryu+2023}). In Figure~\ref{fig:diskprofile}, we present the density, temperature, rotational velocity, and the aspect ratio of the disks in the models considered in Figure~\ref{fig:accretionrate}.

Finally, in Figure~\ref{fig:inclination}, we show the mutual inclination angle between the disk and the binary orbit in models where the final product is a binary consisting of the BH surrounded by a disk. In seven out of ten models considered, the mutual inclination angle between the disk orbit and the binary orbit is not very different from the initial encounter inclination angle, which is not surprising. However, it is quite striking that the final disk-binary orbit inclination angles in the remaining three models (Models 11., 12, and 27) are completely different from the initial encounter inclination angle. Coincidentally they are all retrograde encounters (three out of four). These findings imply that, since in actual astrophysical settings a third body will approach a binary with an arbitrary inclination angle, if a disk forms around a binary member during the three-body interactions, the binary orbit and the disk are not likely aligned at the moment of the disk formation. This may indicate that the orientation of the disk around the BH in BH-star binaries in dense environments can be indicative of the inclination angle of the incoming object in the previous encounter.

\begin{figure}
	\centering

	\includegraphics[width=8.6cm]{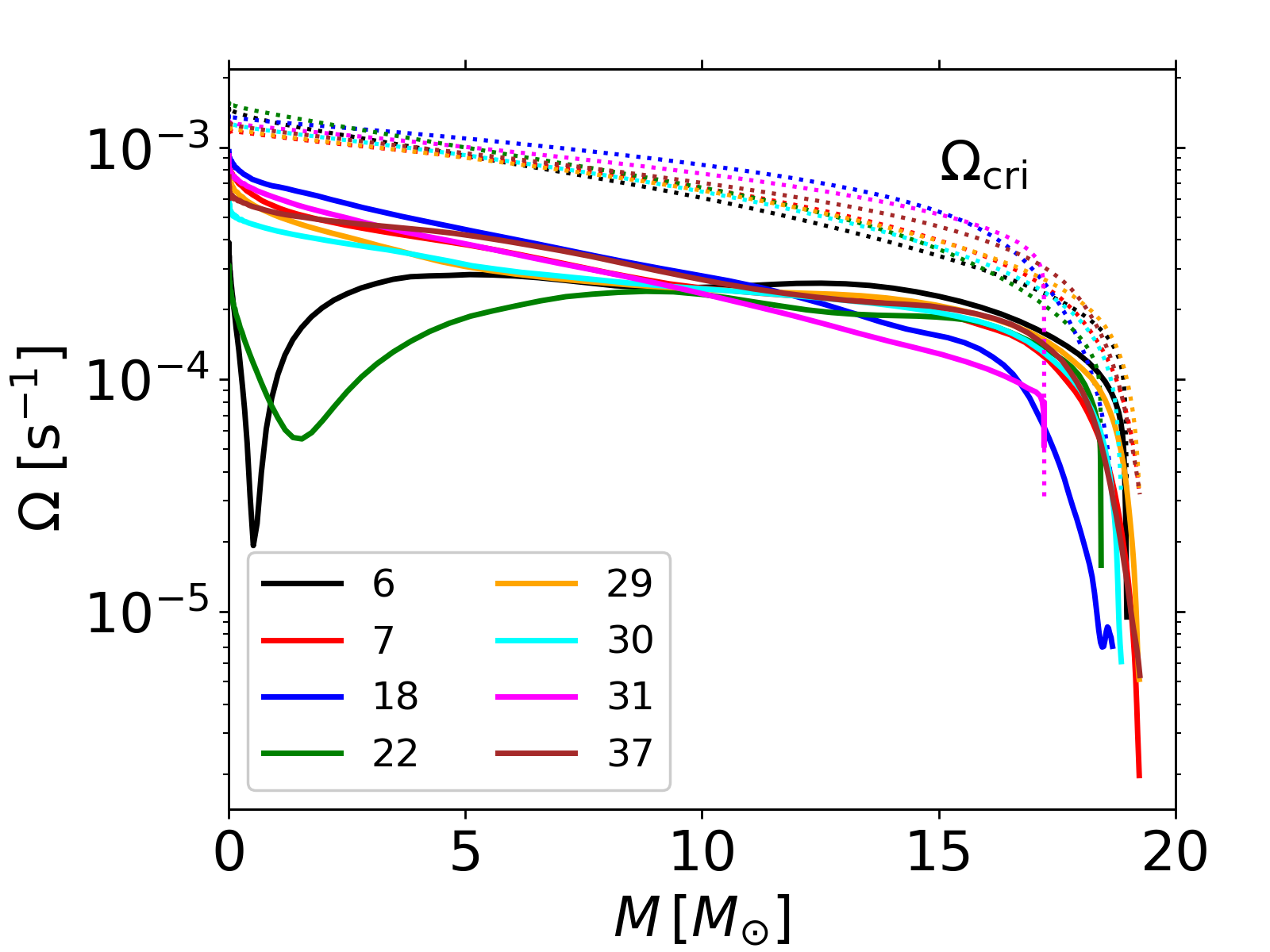}
	\includegraphics[width=8.6cm]{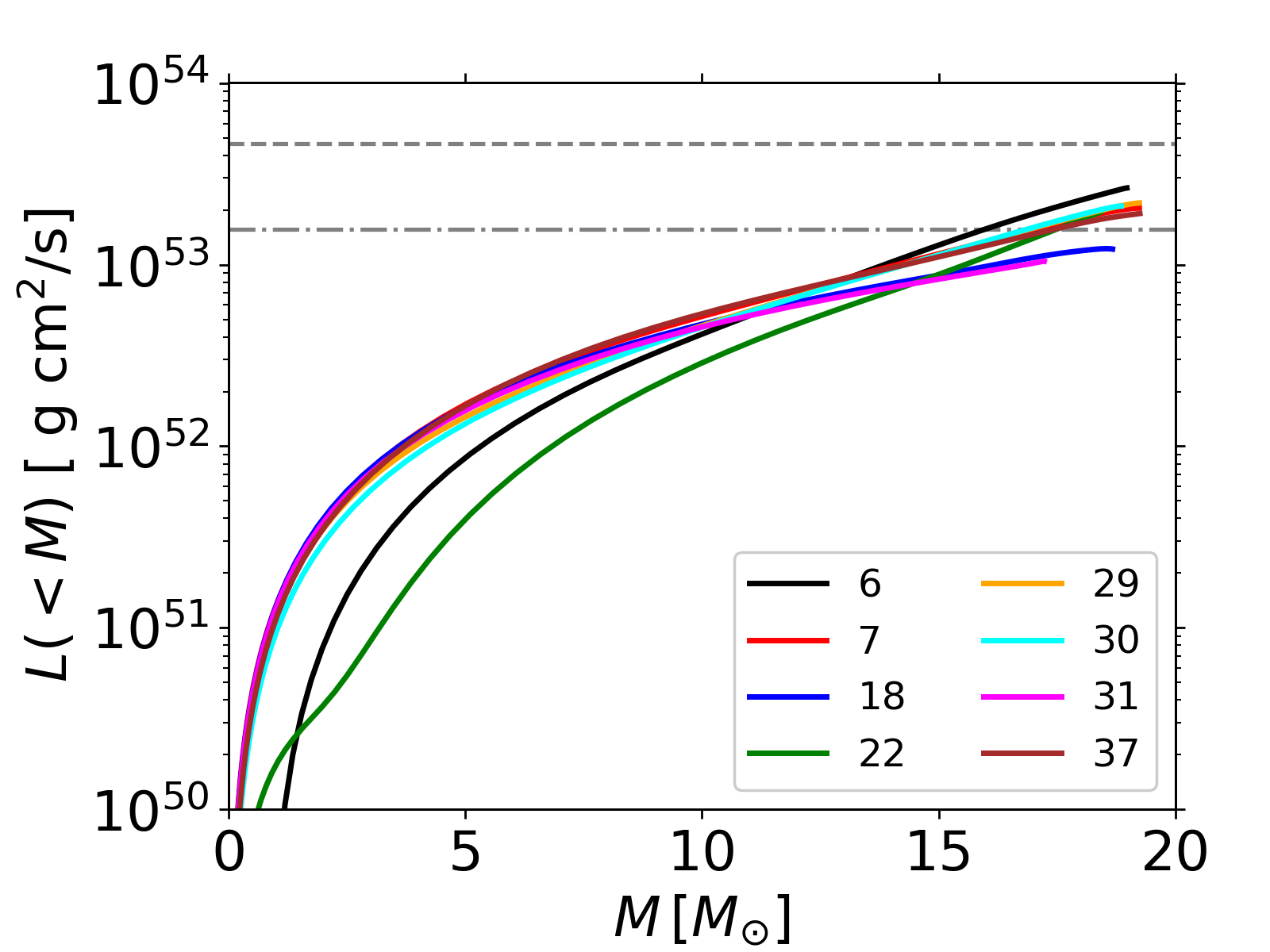}
\caption{Rotational velocity $\Omega$ (\textit{top}) and cumulative angular momentum $L$ (\textit{bottom}) of merged stars (solid lines) as a function of the enclosed mass. The dotted lines in the \textit{upper} panel show the local critical rotational frequency $\Omega_{\rm cri}$, defined as $({G M(<R)/R^{3}})^{1/2}$ where $M(<R)$ is the enclosed mass. The horizontal dashed line in the \textit{bottom} panel show the maximum allowed angular momentum of a MS star with $\mstar\simeq18.5\Msol$ and $\rstar\simeq8.7\Rsol$, and the dot-dashed for the original MS star with $\mstar\simeq10\Msol$ and $\rstar\simeq5.4\Rsol$.}
	\label{fig:merger_L}
\end{figure}

\subsection{Merger product}\label{subsec:merger}

Our simulations show that the two stars in this type of dynamical interactions can merge (12 out of 37 models). The post-merger star often forms a binary with the BH. The center of mass velocity of the binary is typically very low ($3-10 \,{\rm km\,s^{-1}}$).  The mutual orbits of the two stars before the collision are such that $1-e\simeq -0.01 - 0.2$ and the pericenter distance is $(0.05 - 0.5)\times \rstar$, corresponding to the velocity $v_{\rm rel}\simeq 0.7-0.9\, v_{\rm esc}$ at infinity. Here, $v_{\rm esc}=(G\mstar/\rstar)^{1/2} \simeq 600\,{\rm  km \, s^{-1}}$ is the escape velocity of the $10\Msol$ star. In the parameter space considered, mergers almost exclusively occur when the two stars first meet. The fate of the merged stars is diverse. If the merger is able to significantly cancel the momenta of the colliding stars, the merged star is brought on a radial orbit towards the BH and disrupted at pericenter. On the other hand, if the momentum cancellation is not significant, the merged stars can form a binary with the BH. Last is a case (Model 37. $a4b1/2\phi315i30$) in which the incoming star undergoes a close encounter with the BH and  exerts a momentum kick to the BH strong enough to eject it from the two stars. Then the two stars merge, remaining unbound from the BH. 

We find that the dynamically merged stars have a mass of $\sim 18 - 19 \Msol$ after losing $\sim 1 - 2\Msol$ during the merger. This mass loss corresponds to $5 - 10$ percent of the total mass, which is similar to what has been found for equal or similar mass low-velocity ($v_{\rm rel}<v_{\rm esc}$) stellar collisions in previous work \citep[e.g.,][]{Lai+1993,LaycockSills2005,Freitag+2005,DaleDavies2006,Glebbeek+2013}. The thermodynamic state (e.g., density and temperature) of all collision products is not varying significantly among one another. However, collision remnants are significantly puffed up compared to a non-rotating ordinary main sequence (MS) star of the same mass at a similar evolutionary stage (``ordinary'' star), evolved using {\small MESA}\footnote{ Note that we confirmed that the internal structure of the non-rotating star is very similar to that of rotating stars with a rotational speed less than 60\% of their break-up speed, which is roughly the maximum speed of merged stars in our simulations.}, except the one in Model 31 where the merger product has the smallest mass and is substantially more compact than the others. The inflated radii are also similarly found for the coalescence of two stars initially in binaries \citep[e.g.,][]{Schneider+2019}. To demonstrate this, we depict the 1D radially averaged density and temperature profiles in the \textit{top} panels of Figure~\ref{fig:mergerproduct}. As shown in the \textit{top-left} panel, the density profiles of the merger products are not significantly different from each other. However, they are much more extended in size than for the ordinary star (dashed grey). Because of the larger size, the central densities of the merger product ($\sim 5 - 10\,{\rm  g\, cm^{-3}}$) are generally lower than those of the ordinary star; we find them to be lower by a factor of less than two. Similarly, as shown in the \textit{top-right} panel, the overall temperature profiles of the merger products are extended outwards and their temperatures are lower than for the ordinary star.

\begin{figure}
	\centering
	\includegraphics[width=8.4cm] {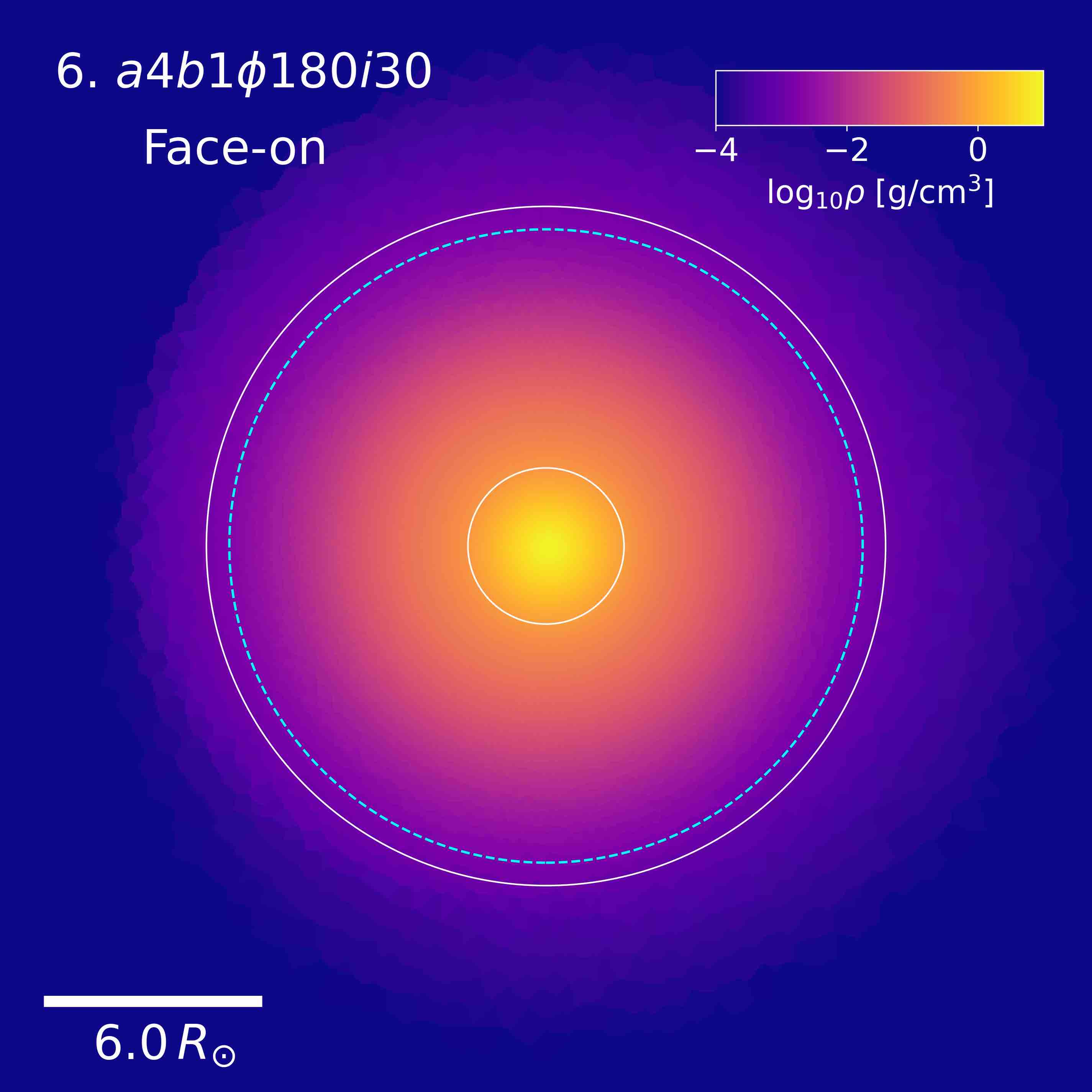}
	\includegraphics[width=8.4cm]{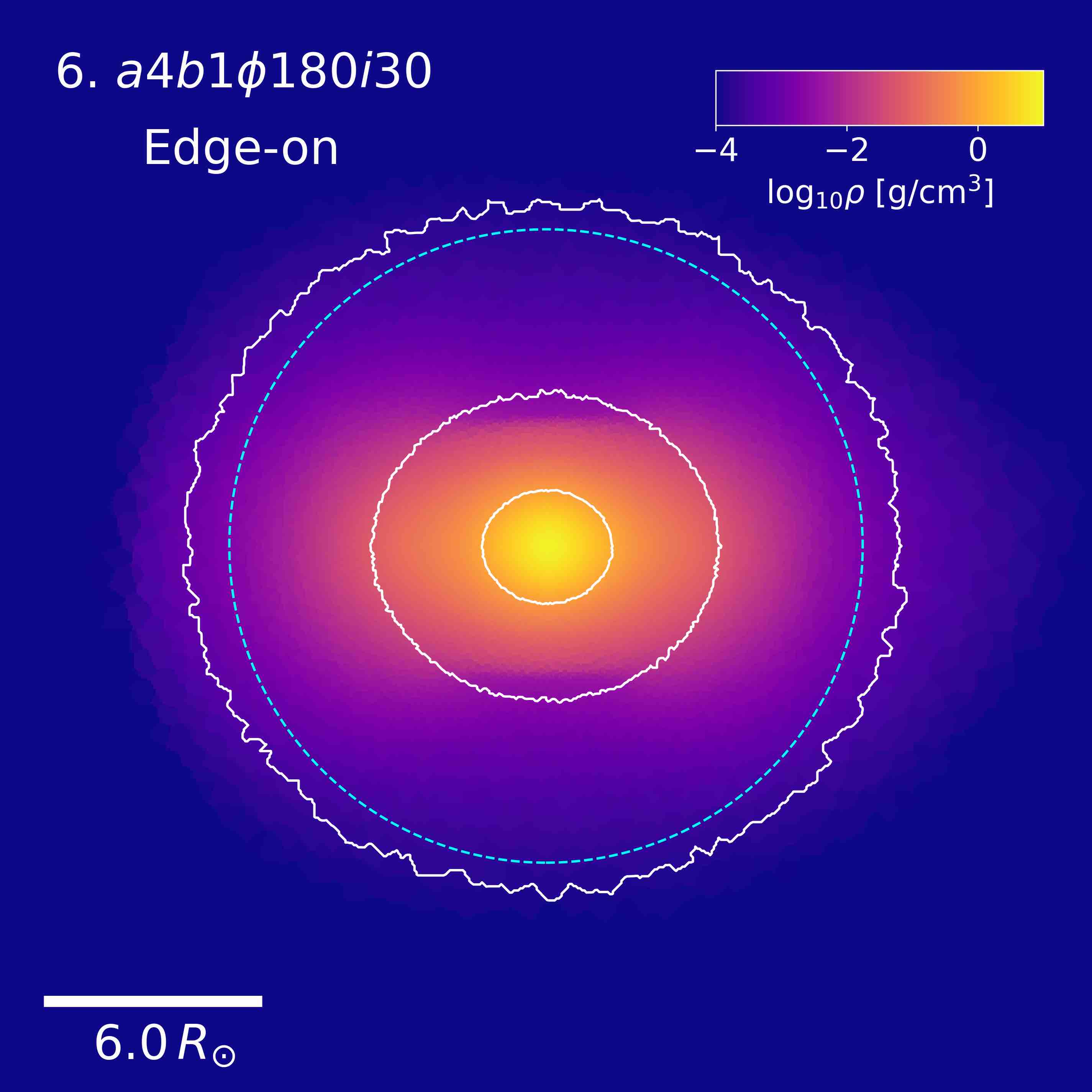}
	\caption{Density (\textit{top}: face-on, \textit{bottom}: edge-on) of a merged star in Model 6. $a4b1/2\phi180i120$ (\textit{top}). The two solid circles in the \textit{top} panel indicate the radius enclosing 50 percent (smaller) and 95 precent (larger) of the remnant mass, respectively, and the white contours in the \textit{bottom} panel give equipotential surfaces. The cyan dashed circle in both panels depicts the radius of an $18.5\Msol$ non-rotating ordinary MS star.}
	\label{fig:mergerproduct2}
\end{figure}

The H ($X_{\rm H}$) and He ($X_{\rm He}$) mass fractions reveal more significant differences from those of the ordinary star, which are shown in the \textit{bottom} panels of Figure~\ref{fig:mergerproduct}. In particular, the core $X_{\rm He}$ of the most merged stars has decreased from 0.68 (initial state) to 0.6. Equivalently, the core $X_{\rm H}$ has increased from 0.3 (initial state) to 0.4. The transition from the core ($X_{\rm He}\simeq 0.6$) to the envelope ($X_{\rm He}\simeq 0.25$) is smoother than in the MESA ordinary non-merger stellar model. As an example, we show how He is mixed in the core during a merger in Model 29. $a4b1/2\phi180i0$ in Figure~\ref{fig:mergerproduct3}. We note that in two cases (Models 6. and 22.), the core $X_{\rm H}\simeq 0.35$, is somewhat smaller than that of most of the merged stars. This difference could originate from a different configuration at collision (e.g., relative speed at collision and impact parameter): less significant mixing ($X_{\rm H}$ closer to its initial value) would have resulted from a collision with a smaller impact parameter, i.e., closer to a head-on collision.  The smaller $X_{\rm He}$ than the initial state indicates that fresh H initially in the envelope of each star is mixed into the merged core during the merger, as similarly shown for unequal-mass stellar collisions in \citet{DaleDavies2006}. Notice that some merger products reveal unstable gradients in the profile, such as an inverted gradient in composition at $R\simeq 15\Rsol$ for Model 6 or in temperature at $R\simeq 12\Rsol$ for Model 22, which likely indicates that the merger products have not reached a fully stable state. However, as the star is settling into a stable state, the inverted gradients will be removed via, e.g., thermohaline mixing of the composition \citep{Kippenhahn+1980,Kippenhahn+1980b}.

The merged stars tend to be differentially rotating, as shown in the \textit{top} panel of Figure~\ref{fig:merger_L}. The rotational frequency $\Omega$ near the core is $6 - 8\times 10^{-4} \sec^{-1}$ and decreases outwards to $\lesssim 10^{-4}\sec^{-1}$ at the surface. This corresponds to $\Omega/\Omega_{\rm cri}\gtrsim 0.4$ within $\simeq 1\Rsol$, and $\Omega/\Omega_{\rm cri}\simeq 0.1 - 0.5$ near the surface. Here $\Omega_{\rm cri}$ is the local critical frequency, defined as $({GM(<R)/R^{3}})^{1/2}$, and $R$ is the distance from the center of mass of the merged star. We also find that the two merger products with relatively low $X_{\rm H}$ (Models 6. and 22.) are rotating more slowly ($\Omega \lesssim 2\times 10^{-4}\sec^{-1}$) near the core, as expected from a head-on collision, and they are closer to rigid rotators than the other merger products. 

Because of the rapid spin, the overall shape of the merged stars takes that of an oblate spheroid. Figure~\ref{fig:mergerproduct2} shows the density in the equatorial plane and a $x-z$ slice of the merger product in Models 6. An interesting remark here is that we do not see any evidence of a disk around the merged stars, as illustrated in the figure. This is consistent with \citet{Sills+2002}. Instead, the star is surrounded by a low-density spherical envelope. Based on their hydrodynamics simulations of off-axis stellar collisions between $0.6\Msol$ and $0.8\Msol$ evolved MS stars, \citet{Sills+2001} posed an ``angular momentum problem'' where merger products are formed with too large angular momentum so that some of the angular momentum has to be lost in order for them to settle into a stable state, possibly blue stragglers. The existence of a disk can mitigate this problem because disk-star interactions, e.g., magnetic locking, can remove the angular momentum of the merger products. Although we do not find a disk surrounding the merger products, their angular momentum is already below critical in our simulations. We show in the \textit{bottom} panel of Figure~\ref{fig:merger_L} the cumulative angular momentum distribution inside the merger products, in comparison with the two maximum angular momentum of the ordinary star (dashed horizontal) and the original $10\Msol$ star (dot-dashed horizontal). The total angular momentum inside the merged stars is $1-3\times10^{53}\,{\rm g\, cm^{2}\, s}$, which is more than a factor of 2 smaller than the maximum angular momentum that the ordinary star with $\mstar \simeq 18.5\Msol$ would have. This means that in principle the merged stars could settle into a stable state without losing any mass due to exceedingly large centrifugal forces.

\section{Discussion}\label{sec:discussion}

\subsection{Electromagnetic Transients}\label{subsec:EMT}

Three-body interactions between BH-star binaries and single stars can create a variety of EMTs. For the parameters considered in this study, four classes can create immediate EMTs. In the class \textit{Stellar disruption}, the stellar debris quickly forms an accretion disk and the BH accretes gas; this, together with shocks, can generate EM radiation. In the class \textit{Merger},  some fraction of mass is ejected at the collision and spread out. Almost instantaneously the BH becomes embedded in a gaseous medium like in a common envelop phase, and can emit radiation via accretion and shocks. Additionally, if the merged star forms a sufficiently compact and highly eccentric binary with the BH (e.g., Model 31), eccentric mass transfer can lead to periodic  EM emission.  Interacting binaries can form also in the last two classes: while we find such a case only in the class \textit{Member exchange} (Model 8), the formation of interacting binaries is in principle possible also in the class \textit{Orbit perturbation}. 
 
 The EM signatures from EMTs are diverse, as illustrated in Figure~\ref{fig:accretionrate}, depending on the encounter configurations and outcomes. To zeroth order, $\dot{M}$ can be a useful proxy for luminosity. In this sense, the various types of $\dot{M}$ and the identification of the encounter types that generate each type of $\dot{M}$ in this work can be used to understand the origin of transients produced in three-body interactions. However, for a more reliable identification of transients, a more systematic investigation covering a wider range of parameters will be required.

\begin{figure}
	\centering

	\includegraphics[width=8.6cm]{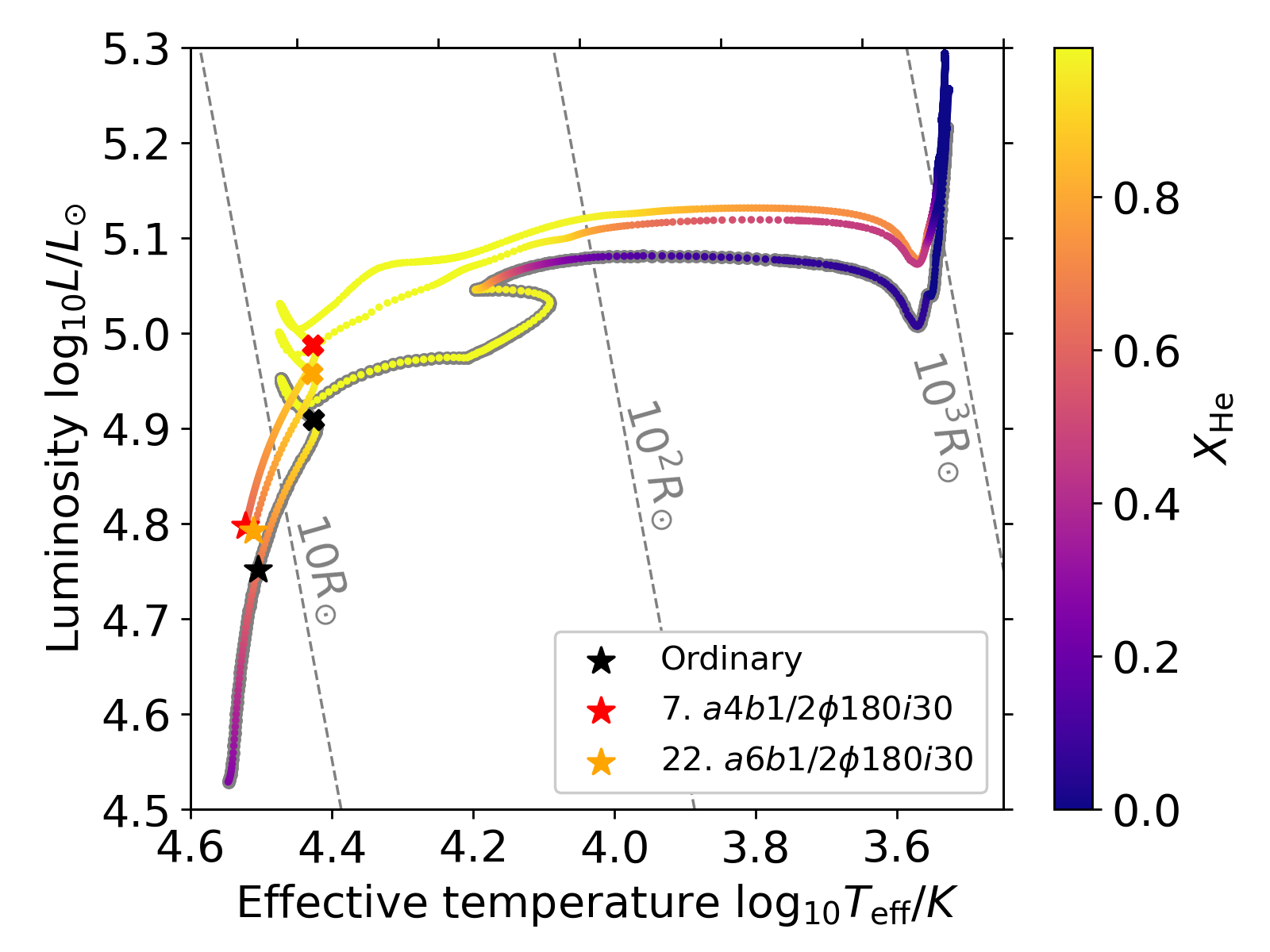}
\caption{Long-term evolution of two merger products (Models 7. $a4b1/2\phi18030$ and 22. $a6b1/2\phi180i30$) using {\small MESA} up to the red supergiant branch in comparison to that of an ordinary non-rotating star with mass $18.5\Msol$ in an Hertzsprung–Russell diagram, color-coded with the central $X_{\rm He}$. The star symbol of each model is indicated at the left end of the line, the age at which the merger products settle, for Models 7. $a4b1/2\phi18030$ (red) and 22. $a6b1/2\phi180i30$ (orange) and at a similar location of the line for the ordinary non-merger case, corresponding to the same $X_{\rm H}$ as the settled merger products. The cross symbols indicate the terminal age main-sequence. The diagonal grey dashed lines depict the $L-T_{\rm eff}$ line for a few given stellar radii. }
	\label{fig:hrdiagram}
\end{figure}

\subsection{Long-term evolution of merger products}

Our simulations show that two stars can collide and merge in three-body interactions, and the merger product can form a binary with the BH or be ejected from the BH. The rate of such stellar collisions in three-body interactions involving binaries can be significantly large compared to that between two single stars in clusters, due to mass segregation and a large encounter cross section \citep{PortegiesZwart+1999}.  We also showed that the internal structure of the merger products, after being dynamically settled, is different from that of an ordinary star of the same mass and metallicity at a similar age that has not undergone any merger (Figure~\ref{fig:mergerproduct}). First, the merger products have larger radii than those of ordinary stars (by almost a factor of $2-3$), indicating that the merger products are not in thermal equilibrium. Second, the core hydrogen fraction can be enhanced by 30 percent compared to that of the original star before merger. Equivalently, the core helium fraction can be lower by a similar amount. Lastly, the merged stars tend to be differentially rotating at $0.1-0.5$ of the critical rotational velocity. All of these properties are qualitatively very similar to those of partially disrupted stars \citep{Ryu+2020c, Ryu+2023}.  We note that magnetic fields, if included, can be significantly enhanced in merger products \citep[e.g.,][]{Schneider+2019,Schneider+2020}.

Given the peculiarity of the merger products, we investigate their long-term evolution  using \mesa. We create a non-rotating zero-age main-sequence star with the same metallicity as the original $10\Msol$ star (i.e., $Z=0.006$). Then we relax the star until its entropy, mass and chemical composition distribution match those of the merger product. This is achieved by iteratively modifying the normal stellar model over 1000 steps under the condition that the internal structure satisfies the stellar structure equations. Then we evolve the relaxed star using the wind and overshoot prescriptions adopted to create the original $10\Msol$ star. In this analysis, we ignore rotation. Figure~\ref{fig:hrdiagram} shows the evolution of two models (Models 7. $a4b1/2\phi180i30$ and 22. $a6b1/2\phi180i30$) for the next 5-6 million years since merger, and that of the ordinary star with mass of $18.5\Msol$, in a Hertzsprung-Russell diagram. The core helium fraction for the merger products is always higher than that of the ordinary star at similar locations in the diagram. The evolutionary tracks of the merger products are generally located above the track of the ordinary star, implying that the merger products are hotter and more luminous at any given stellar age. This is qualitatively consistent with previous work on MS stellar mergers \citep[e.g., collisions between middle-age MS stars with similar masses][]{Glebbeek+2013}. However, their temperature and luminosity are not significantly larger, at most by a factor of $1.3$. 

We should note that it would be important to include rotation in this analysis given rotation-induced mixing \citep{MeynetMaeder1997}. Although the total angular momentum of the merger products is smaller than the critical value, the merger products can still lose mass due to spin depending on the angular momentum distribution inside the star \citep{Heger+2000}. If the core with mass $M_{\rm core}$ retains an angular momentum larger than that at the innermost stable circular orbit, $j>GM_{\rm core}/c\simeq 2\times 10^{6}\, (M_{\rm core}/2\Msol)\, {\rm cm^{2}\, s^{-1}}$ \citep{Podsiadlowski+2004}, by the time the core collapses, the merger products can become progenitors of hypernovae and long-duration gamma ray bursts. All of this implies that the evolutionary tracks, when the spin is taken into account, could be different from the tracks shown above and have unique astrophysical implications. Given such potential effects on the evolution, we will examine the impact of rotation on the long-term evolution of merger products with proper modeling of rotation and resulting mass loss in future work.

\subsection{Encounters with different masses}\label{subsec:differentmass}

In this study, we consider two stars of the same mass in the three-body interactions and the mass ratio of the stars to the BH is fixed at $0.5$. However, in  realistic cluster environments, the mass of the two stars are not necessarily the same. Also, the star-BH
 mass ratio would be variable. Nonetheless, many of our findings can still apply to this type of three-body encounters with varying masses. If the impact parameter $b$ is less than $a$, interactions would still possibly become violent, independently of the mass ratio. In addition, whether outcomes are mergers between two stars or disruption of a star(s) by the BH would be primarily determined by which two objects meet. The final outcome types, their properties, and their formation frequency would depend on the mass of the incoming star and its mass ratio to the binary mass, like other encounter parameters. For example, if a smaller intruder can play a role as a catalyst for violent interactions \citep[e.g., mergers,][]{Gaburov+2010}, stellar mergers, TDEs, and star-BH collisions would be more frequent. However, if the incoming mass is too small compared to the masses of both binary members, the immediate impact of close encounters (e.g., dissociation of the binary) would be relatively small. For this case, if a merger occurs between two unequal mass stars, the internal structure of the merger product and its long-term evolution could be significantly different from what we found for equal-mass collisions, which would probably result in the strongest mixing \citep[see][]{Glebbeek+2013}. On the other hand, if the incoming star is much more massive than the mass of the star in the binary, then the encounters would be effectively a two-body problem between the incoming star and the BH.

\subsection{Runaway star and black holes}
We showed that this type of three-body encounters can create single stars ejected at velocities of $120 - 240\,{\rm km\, s^{-1}}$, much greater than the typical escape speed of globular clusters \citep[``runaway stars''][]{Blaauw1961,Stone1979,Sana+2022}, as well as single BHs also ejected at high velocities of $63 - 115\,{\rm km\, s^{-1}}$. In particular, some of the rapidly moving BHs had undergone a TDE and became surrounded by an accretion disk, meaning they are emitting radiation while being ejected. If the lifetime of the accretion disk around them is sufficiently long, those could be observed as rapidly moving runaway BHs outside clusters.

\subsection{Encounter rate in globular clusters}\label{subsec:rate}

Following \citetalias{Ryu+2023} and \citetalias{Ryu+2023b}, we first make an order-of-magnitude estimate for the differential rate of a BH-star binary encountering a single star per single star as ${\rm d}\mathcal{R}/{\rm d} N_{\rm s}\simeq n\Sigma v_{\rm rel}$. Here, $n$ is the binary number density near the cluster center, $v_{\rm rel}$ the relative velocity between the binary and the single star, and $\Sigma$ the encounter cross-section. We adopt the estimate for ${\rm d}\mathcal{R}/{\rm d} N_{\rm s}$ made in \citetalias{Ryu+2023b}, 
\begin{align}\label{eq:rate}
   \frac{{\rm d}\mathcal{R}}{{\rm d} N_{\rm s}} 
              &\simeq 4 \times 10^{-14} \yr^{-1} \left(\frac{f_{\rm b}}{10^{-5}}\right) \left(\frac{n_{\rm s}}{10^{5}{\rm pc}^{-3}}\right)\left(\frac{M_{\bullet} +\mstar}{20\Msol}\right)\nonumber\\
              &\times\left(\frac{a}{100\Rsol}\right) \left(\frac{\sigma}{15\km\sec}\right)^{-1},
\end{align}
where we express $n$ as $n \simeq f_{\rm b}n_{\rm s}$, $f_{\rm b}$ is the non-interacting star - BH binary fraction $\simeq 10^{-4}-10^{-5}$ \citep{Morscher+2015}, $n_{\rm s}$ gives the number density of stellar-mass objects, and $\sigma$ is the velocity dispersion. Because the number of single stars in the core of size $r_{\rm c}\simeq 1\pc$ is $N_{\rm s}\simeq 4\pi r_{\rm c}^{3} n_{\rm s}/3 \simeq 4\times 10^{5}$, the rate of strong three-body encounters per globular cluster is, 
\begin{align}\label{eq:rate1}
   \mathcal{R}& \simeq 2\times 10^{-8} \yr^{-1} \left(\frac{r_{\rm c}}{1\pc}\right)^{3}\left(\frac{f_{\rm b}}{10^{-5}}\right) \left(\frac{n_{\rm s}}{10^{5}{\rm pc}^{-3}}\right)^{2}\left(\frac{M_{\bullet} + \mstar}{20\Msol}\right)\nonumber\\
              &\times\left(\frac{a}{100\Rsol}\right) \left(\frac{\sigma}{15\km\sec}\right)^{-1}.
\end{align}
Assuming $\simeq$150 globular clusters in the Milky Way \citep{Harris+2010}, $\mathcal{R}\simeq3\times 10^{-6}$ per year per galaxy. As noted in \citetalias{Ryu+2023b}, a more precise estimate of $\mathcal{R}$ requires a more careful consideration of cluster evolution history.

\section{Conclusions}\label{sec:conclusion}

Multi-body dynamical interactions, a fundamental mechanism responsible for the evolution of star clusters, have been studied mostly using $N-$body simulations even though non-hydrodynamical effects are essential for determining outcomes and their observables. Continuing our efforts of bringing our understanding of three-body interactions beyond the point-particle approximation, we have investigated the outcomes of three-body encounters between a $20\Msol$ BH -- $10\Msol$ star circular binary and a $10\Msol$ star, using a suite of hydrodynamical simulations with the moving-mesh code {\small AREPO}, for a wide range of encounter parameters.

The results of our simulations are summarized in the following.
\begin{enumerate}
    \item \textit{Three-body encounters between BH-star binaries and single stars can produce five different outcomes, stellar disruption, merger, orbit perturbation, member exchange, and triple formation}. Although in principle the essence if these outcomes can be identified with the point particle approximation assuming finite sizes of the mass points, we obtained the properties of the outcomes and their observables in detail, which can not be studied with $N-$body simulations alone, see Section~\ref{sub:class} for detailed properties of them, such as their plausible formation configurations. 

    \item \textit{The phase angle and the impact parameter play the most important role in determining the outcomes}, similarly to the three-body interactions between star-BH binaries and single BHs studied in \citetalias{Ryu+2023b}. The phase angle determines which two objects first meet: if two stars meet first, one likely outcome is a stellar merger. whereas if the incoming star and the BH interact closely first, the star is destroyed in a tidal disruption event or a collision with the BH. The impact parameter sets the zeroth-order boundary between violent, star-destroying events ($b < 1$) and non-violent events ($b > 1$). However, even for $b<1$, the outcomes can vary depending on the phase angle. 
The probability of having disruptive events is further enhanced when the encounter is in a prograde direction in which the encounter cross section is large because of smaller relative velocities. 

\item \textit{The accretion rate produced in stellar disruptions is mostly super-Eddington and displays various shapes}, depending on the configuration at disruption (e.g., single full disruption, partial disruption, collision, or multiple disruptions, see Figure~\ref{fig:accretionrate}). Accretion timescales are generally a few to ten days, comparable to the duration of fast blue optical transients.

\item \textit{The merger products are hotter and larger than an ordinary star of the same mass at a similar age, and are rotating at 30-50 percent of the critical value.} Those stars stay hotter and brighter than the ordinary star for the next 5 - 6 million years until they become red supergiants. 

\end{enumerate}

We considered similar encounter parameters in this study as those in \citetalias{Ryu+2023b}. The only difference is the type of the incoming object: a star in this study and a BH in \citetalias{Ryu+2023b}, while the masses of the incoming object and the parameters of the original binary are the same. Nonetheless, the two types of three-body encounters produce substantially different types of final outcomes and properties. Importantly, the stellar mergers found in the present study can have important implications for the subsequent long-term evolution of binaries consisting of a merger product formed dynamically in clusters (see Figure~\ref{fig:hrdiagram}). Although our simulations cover a wide range of encounter parameters, the entire parameter space of the three-body interactions remains vast. Nonetheless, out investigation has identified key outcomes such as tidal disruption events and stellar mergers, leaving larger or more focused parametric studies to future explorations. In addition, we will investigate the impact of magnetic fields and background gas on the outcomes and their observables in our future work.

\section*{Acknowledgements}
TR is grateful to Stephen Justham and Earl Bellinger for fruitful discussions of stellar mergers and the evolution of the merger products. This research project was conducted using computational resources (and/or scientific computing services) at the Max-Planck Computing \& Data Facility. Some of the simulations were performed on the national supercomputer Hawk at the High Performance Computing Center Stuttgart (HLRS) under the grant number 44232. The authors gratefully acknowledge the scientific support and HPC resources provided by the Erlangen National High Performance Computing Center (NHR@FAU) of the Friedrich-Alexander-Universität Erlangen-Nürnberg (FAU) under the NHR project b166ea10. NHR funding is provided by federal and Bavarian state authorities. NHR@FAU hardware is partially funded by the German Research Foundation (DFG) – 440719683. The authors would like to also thank Stony Brook Research Computing and Cyberinfrastructure, and the Institute for Advanced Computational Science at Stony Brook University for access to the high-performance SeaWulf computing system, which was made possible by a \$1.4M National Science Foundation grant (\#1531492).
R. Perna acknowledges support by NSF award AST-2006839.

\section*{Data Availability}
Any data used in this analysis are available on reasonable request from the first author.

\appendix

\renewcommand{\thefigure}{A1}

\begin{figure*}
	\centering
		\includegraphics[width=8.6cm]{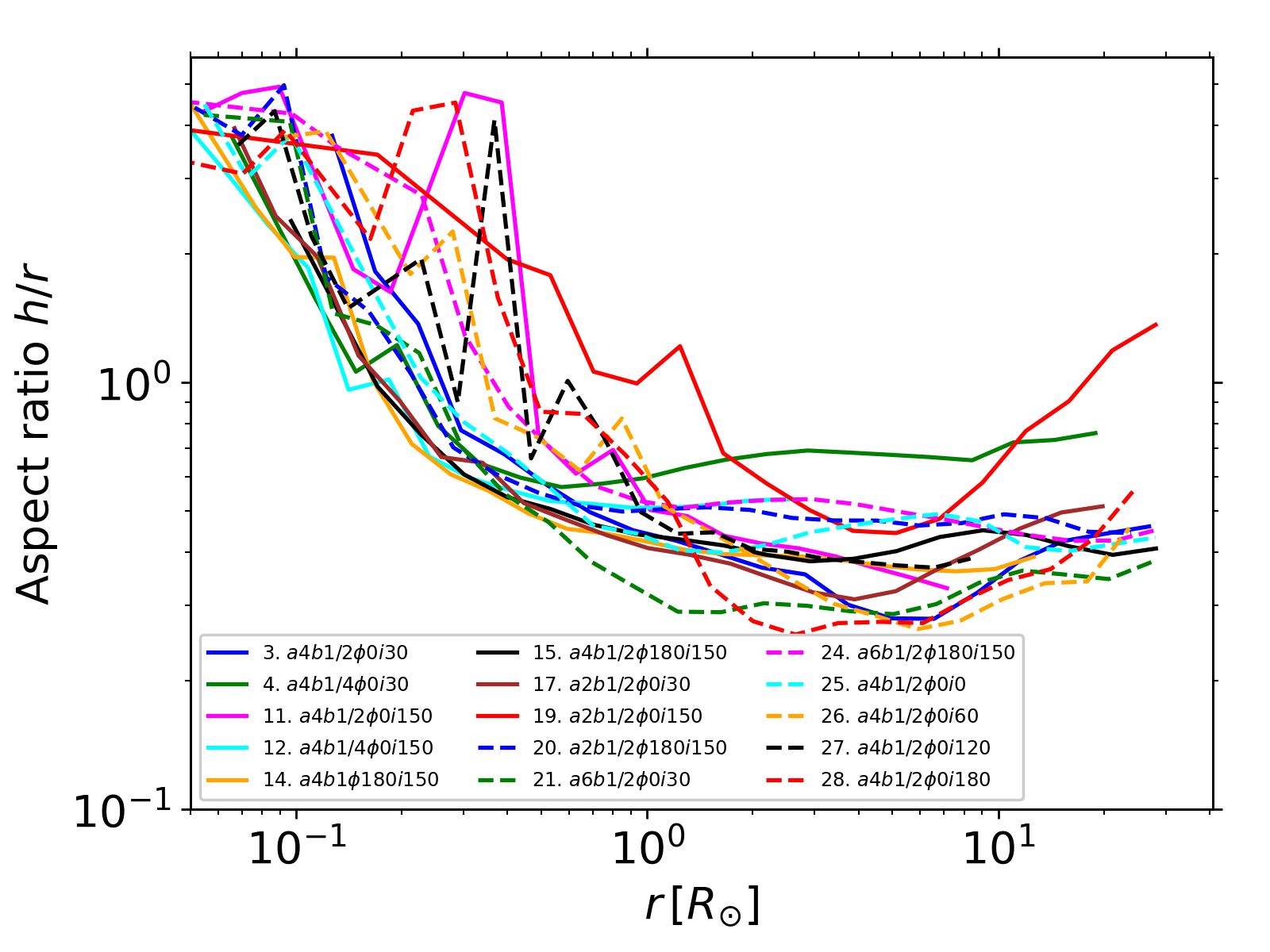}
		\includegraphics[width=8.6cm]{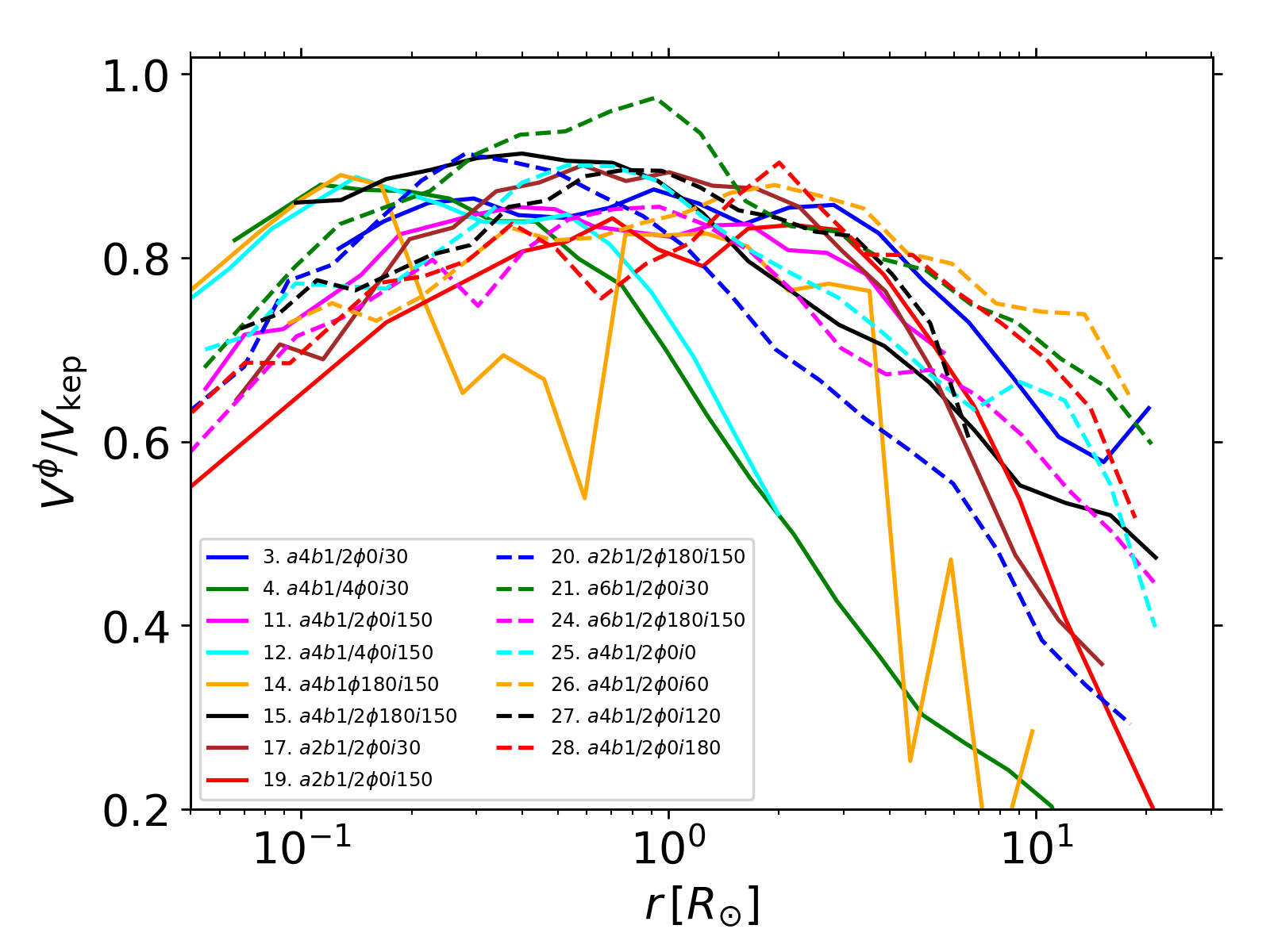}\\
	\includegraphics[width=8.6cm]{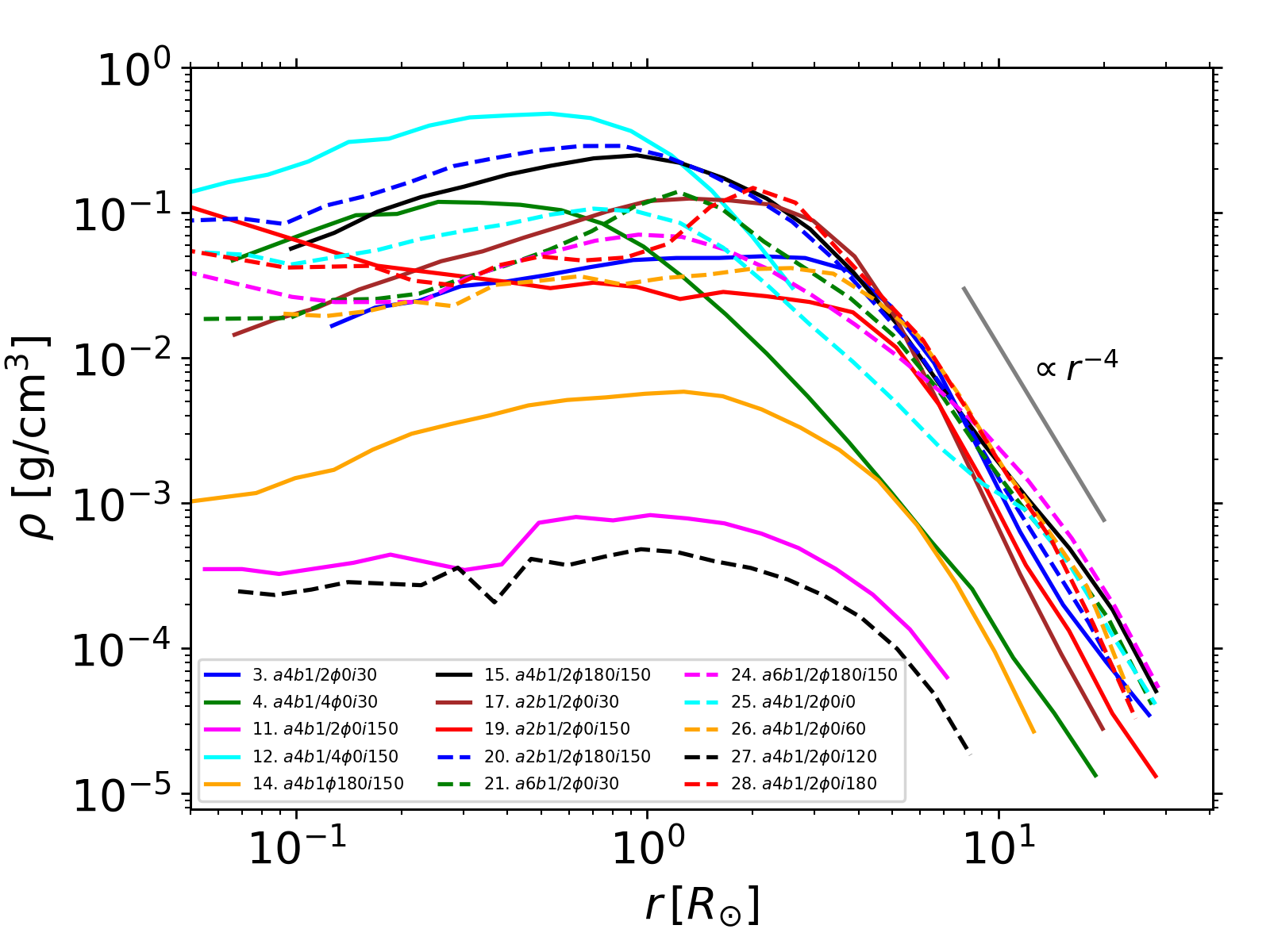}
	\includegraphics[width=8.6cm]{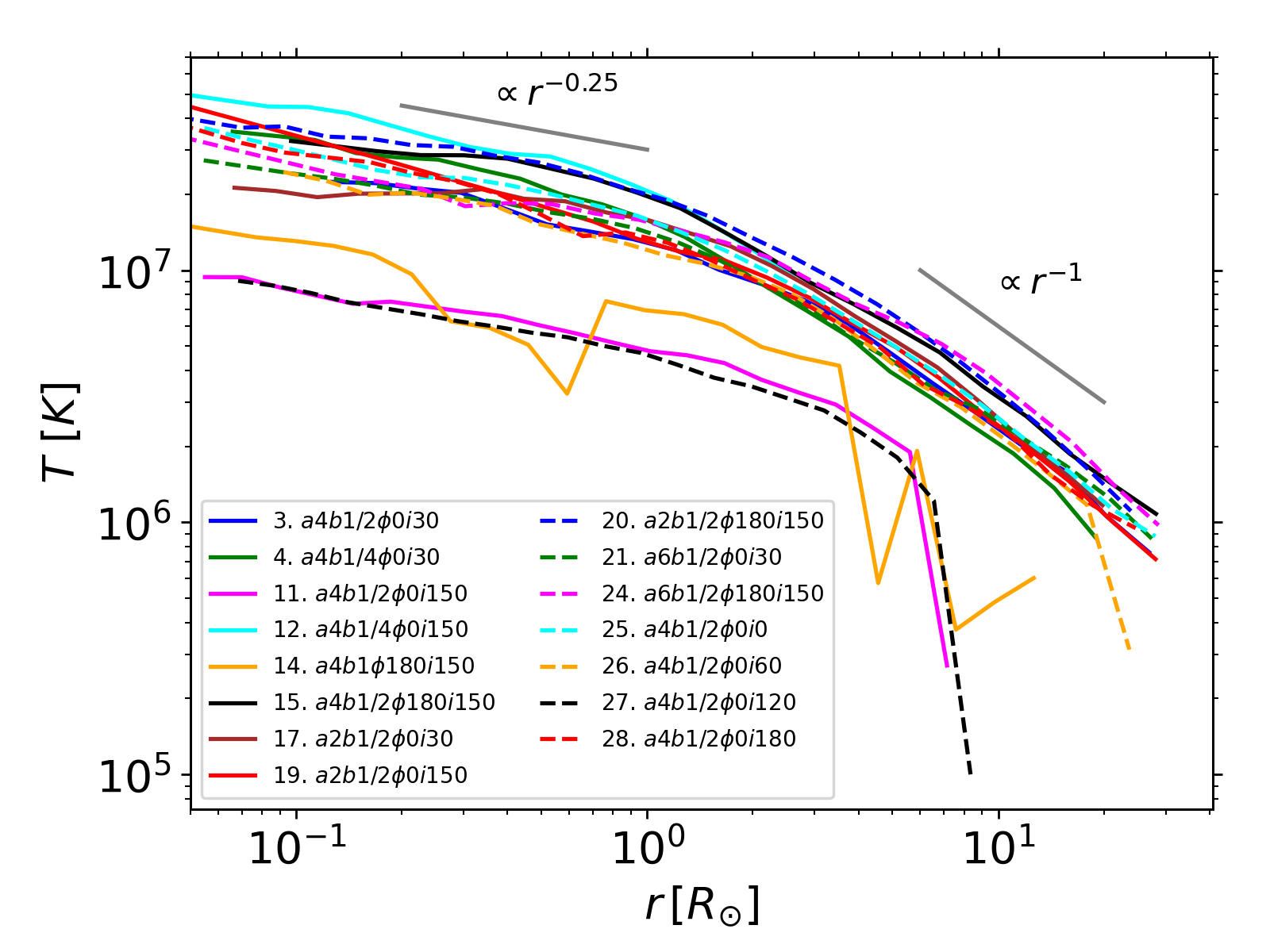}
\caption{Profiles of the structure of the disks measured at the end of our simulations: (\textit{top-left}) the aspect ratio, defined as the ratio of the density scale height to the cylindrical radius $r$, (\textit{top-right}) the ratio of the mass-weighted average of the azimuthal velocity along the midplane within the scale height to the Keplerian velocity $v_{\rm kep}$, (\textit{bottom-left}) the average density along the midplane within the scale height, and (\textit{bottom-right}) the mass-weighted average of the temperature along the midplane within the scale height. }
	\label{fig:diskprofile}
\end{figure*}

\bibliographystyle{mnras}

\section{Disk properties}

We provide the profiles of the aspect ratio (\textit{top-left}), the ratio of the azimuthal velocity to the Keplerian velocity (\textit{top-right}), density (\textit{bottom-left}), and temperature (\textit{bottom-right}) of disks produced during dynamical interactions in Figure~\ref{fig:diskprofile}.

\end{document}